\documentclass[aps,prx,twocolumn,superscriptaddress,notitlepage,floatfix]{revtex4-1}
\usepackage{amsmath,amsfonts,amssymb,color,epsfig,graphics,graphicx,latexsym,theorem,url,multirow,diagbox}
\usepackage[colorlinks,colorlinks,citecolor=blue,linkcolor=magenta,urlcolor=blue]{hyperref}
\usepackage[utf8]{inputenc}
\usepackage{booktabs}
\usepackage[T1]{fontenc}
\usepackage{courier}
\usepackage{listings}
\usepackage{dcolumn}
\usepackage{comment}
\usepackage{times}

\usepackage{algorithm}  
\usepackage{algorithmicx}  
\usepackage{algpseudocode}  
\usepackage{bm}

\DeclareMathOperator{\sign}{sign}
\DeclareMathOperator*{\argmin}{argmin}
\DeclareMathOperator*{\argmax}{argmax}

\begin{document}
\begin{abstract}
Adversarial machine learning is an emerging field that focuses on studying vulnerabilities of machine learning approaches in adversarial settings and developing techniques accordingly to make learning robust to adversarial manipulations. It plays a vital role in various machine learning applications and has attracted tremendous attention across different communities recently. In this paper, we explore different adversarial scenarios in the context of quantum machine learning. We find that, similar to traditional classifiers based on classical neural networks, quantum learning systems are likewise vulnerable to crafted adversarial examples, independent of whether the input data is classical or quantum. In particular, we find that a quantum classifier that achieves nearly the state-of-the-art accuracy can be conclusively deceived by adversarial examples obtained via adding imperceptible perturbations to the original legitimate samples. This is explicitly demonstrated with quantum adversarial learning in different scenarios, including classifying real-life images (e.g., handwritten digit images in the dataset MNIST), learning phases of matter (such as, ferromagnetic/paramagnetic orders and symmetry protected topological phases), and classifying quantum data. Furthermore, we show that based on the information of the adversarial examples at hand, practical defense strategies can be designed to fight against a number of different attacks. Our results uncover the notable vulnerability of quantum machine learning systems to adversarial perturbations, which not only reveals a novel perspective in bridging machine learning and quantum physics in theory but also provides valuable guidance for practical applications of quantum classifiers based on both near-term and future quantum technologies.
\end{abstract}

\title{Quantum Adversarial Machine Learning}

\author{Sirui Lu}
\affiliation{Center for Quantum Information, IIIS, Tsinghua University, Beijing 100084, People\textquoteright s Republic of China}
\affiliation{Max-Planck-Institut f\"ur Quantenoptik, Hans-Kopfermann-Str.\ 1, D-85748 Garching, Germany}
\author{Lu-Ming Duan}\email{lmduan@tsinghua.edu.cn}
\affiliation{Center for Quantum Information, IIIS, Tsinghua University, Beijing 100084, People\textquoteright s Republic of China}
\author{Dong-Ling Deng}\email{dldeng@tsinghua.edu.cn}
\affiliation{Center for Quantum Information, IIIS, Tsinghua University, Beijing 100084, People\textquoteright s Republic of China}
\date{\today}

\maketitle

\section{Introduction}\label{introduction}

\begin{figure}
\centering
\includegraphics[width=0.48\textwidth]{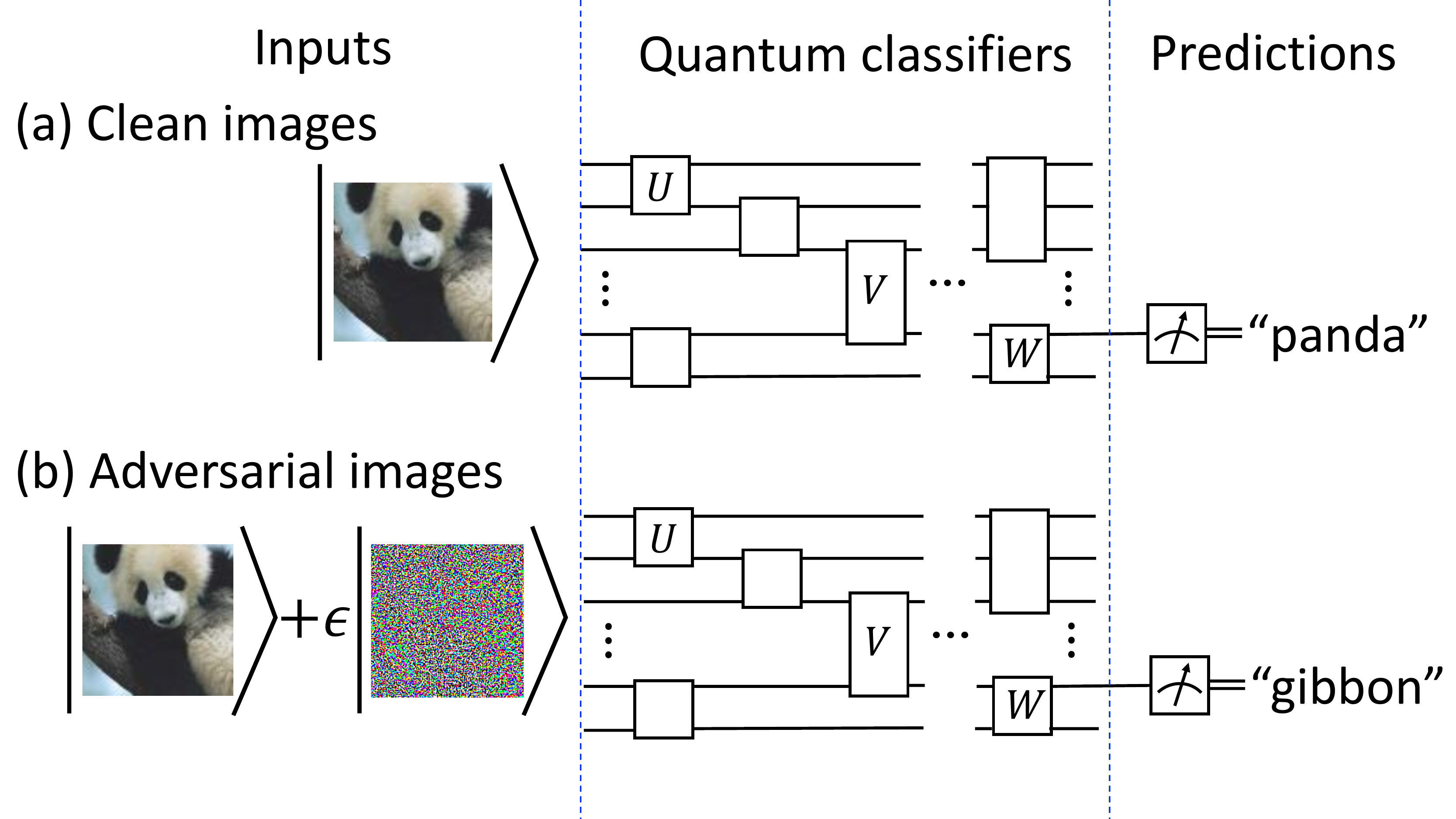}
\caption{\label{fig:QuantumPanda} A schematic illustration of quantum adversarial machine learning. (a) A quantum classifier that can successfully identify the image of a panda as ``panda'' with the state-of-the-art accuracy. (b) Adding a small amount of carefully-crafted noise will cause the same quantum classifier to misclassify the slightly modified image, which is indistinguishable from the original one to human eyes, into a ``gibbon'' with notable high confidence.}
\end{figure}

The interplay between machine learning and quantum physics may lead to unprecedented perspectives for both fields \cite{Sarma2019Machine}. On the one hand, machine learning, or more broadly artificial intelligence, has progressed dramatically over the past two decades \cite{Jordan2015Machine, Lecun2015Deep} and many problems that were extremely challenging or even inaccessible to automated learning have been solved successfully \cite{Silver2016Mastering,Silver2017Mastering}.
This raises new possibilities for utilizing machine learning to crack outstanding problems in quantum science as well \cite{Sarma2019Machine,Carleo2016Solving,Torlai2018Neural,Chng2017Machine,Nomura2017Restricted,Wang2016Discovering,You2017Machine,Deng2017Machine,Deng2017MachineBN,Deng2017Quantum,Gao2017Efficient,melko2019restricted}. On the other hand, the idea of quantum computing has revolutionized theories and implementations of computation, giving rise to new striking opportunities to enhance, speed up or innovate machine learning with quantum devices, in turn \cite{Biamonte2017Quantum,Dunjko2018Machine, Ciliberto2017Quantum}. This emergent field is growing rapidly, and notable progress is made on a daily basis. Yet, it is largely still in its infancy, and many important issues remain barely explored \cite{Sarma2019Machine,Biamonte2017Quantum,Dunjko2018Machine, Ciliberto2017Quantum}. In this paper, we study such an issue concerning quantum machine learning in various adversarial scenarios. We show, with concrete examples, that quantum machine learning systems are likewise vulnerable to adversarial perturbations (see Fig.~\ref{fig:QuantumPanda} for an illustration) and suitable countermeasures should be designed to mitigate the threat associated with them.

In classical machine learning, the vulnerability of machine learning to intentionally-crafted adversarial examples as well as the design of proper defense strategies has been actively investigated, giving rise to an emergent field of adversarial machine learning \cite{Huang2011Adversarial,szegedy2013intriguing,goodfellow2014explaining,biggio2018wild,miller2019adversarial,vorobeychik2018adversarial,carlini2017adversarial,BIM,ilyas2018black,tjeng2018evaluating,athalye2018obfuscated,papernot2017practical,madry2017towards,chen2017zoo}.
Adversarial examples are inputs to machine learning models that an attacker has crafted to cause the model to make a mistake.
The first seminal adversarial example dates back to 2004 when Dalvi \emph{et al.} studied the techniques used by spammers to circumvent spam filters \cite{Dalvi2004Adversarial}. It was shown that linear classifiers could be easily fooled by few carefully-crafted modifications (such as adding innocent text or substituting synonyms for words that are common in malignant message) in the content of the spam emails, with no significant change of the meaning and readability of the spam message. Since then, adversarial learning has attracted enormous attention, and different attack and defense strategies were proposed \cite{madry2017towards,BIM,MIM,CW,goodfellow2014explaining,chen2017zoo}.
More strikingly, adversarial examples can even come in the form of imperceptibly small perturbations to input data, such as making a human-invisible change to every pixel in an image \cite{szegedy2013intriguing,nguyen2015deep,moosavi2016deepfool}. A prominent example of this kind in the context of deep learning was first observed by Szegedy \emph{et al.} and has been nowadays a celebrated prototype example that showcases the vulnerability of machine learning in a dramatic way \cite{szegedy2013intriguing}: starting with an image of a panda, an attacker may add a tiny amount of carefully-crafted noise (which is imperceptible to the human eye) to make the image be classified incorrectly as a gibbon with notably high confidence. In fact, the existence of adversarial examples is now widely believed to be ubiquitous in classical machine learning. Almost all type of learning models suffer from adversarial attacks, for a wide range of data types including images, audio, text, and other inputs \cite{biggio2018wild,miller2019adversarial}. From a more theoretical computer science perspective, the vulnerability of classical classifiers to adversarial perturbations is reminiscent of the ``No Free Lunch'' theorem---there exists an intrinsic tension between adversarial robustness and generalization accuracy \cite{tsipras2018robustness, fawzi2018adversarial, gilmer2018adversarial}. More precisely, it has been proved recently that if the data distribution satisfies the $\text{W}_2$ Talagrand transportation-cost inequality (a general condition satisfied in a large number of situations, such as the cases where the class-conditional distribution has log-concave density or is the uniform measure on a compact Riemannian manifold with positive Ricci curvature), any classical classifier could be adversarially deceived with high probability \cite{dohmatob2019generalized}.

Meanwhile, over the past few years, a number of intriguing quantum learning algorithms have been discovered \cite{Biamonte2017Quantum,Lloyd2018Quantum,Demers2018Quantum,schuld2018circuit,Zeng2019Learning,farhi2018classification,schuld2017implementing,mitarai2018quantum,schuld2019quantum,havlivcek2019supervised,zhu2018training,cong2018quantum,wan2017quantum,grant2018hierarchical,du2018implementable,uvarov2019machine,Gao2017anEfficient,liu2018differentiable,perdomo2018opportunities,amin2016quantum}, and some been demonstrated in proof-of-principle experiments \cite{hu2019quantum}. These algorithms exploit the unique enigmatic properties of quantum phenomena (such as superposition and entanglement) and promise to have exponential advantages compared to their classical counterparts. Notable examples include the HHL (Harrow-Hassidim-Lloyd) algorithm \cite{Harrow2009Quantum}, quantum principal component analysis \cite{cong2016quantum}, quantum support-vector machine \cite{rebentrost2014quantum,Li2015Experimental}, and quantum generative model \cite{Gao2017anEfficient}, etc. Despite this remarkable progress, quantum learning within different adversarial scenarios remains largely unexplored \cite{liu2019vulnerability,wiebe2018hardening}. A noteworthy step along this direction has been made recently by Liu and Wittek \cite{liu2019vulnerability}, where they showed in theory that a perturbation by an amount scaling inversely with the Hilbert space dimension of a quantum system to be classified should be sufficient to cause a misclassification, indicating a fundamental trade-off between the robustness of the classification algorithms against adversarial attacks and the potential quantum advantages we expect for high-dimensional problems. Yet, in practice, it is unclear how to obtain adversarial examples in a quantum learning system, and the corresponding defense strategy is lacking as well.

In this paper, we study the vulnerability of quantum machine learning to various adversarial attacks, with a focus on a specific learning model called quantum classifiers. We show that, similar to traditional classifiers based on classical neural networks, quantum classifiers are likewise vulnerable to carefully-crafted adversarial examples, which are obtained by adding imperceptible perturbations to the legitimate input data. We carry out extensive numerical simulations for several concrete examples, which cover different scenarios with diverse types of data (including handwritten digit images in the dataset MNIST, simulated time-of-flight images in cold-atom experiment, and quantum data from an one-dimensional transverse field Ising model) and different attack strategies (such as, fast gradient sign method \cite{madry2017towards}, basic iterative method \cite{BIM}, momentum iterative method \cite{MIM}, and projected gradient descent \cite{madry2017towards} in the white-box attack setting, and transfer-attack method \cite{papernot2016transferability} and zeroth-order optimization \cite{chen2017zoo} in the black-box attack setting, etc.) to obtain the adversarial perturbations. Based on these adversarial examples, practical defense strategies, such as adversarial training, can be developed to fight against the corresponding attacks. We demonstrate that, after the adversarial training, the robustness of the quantum classifier to the specific attack will increase significantly. Our results shed new light on the fledgling field of quantum machine learning by uncovering the vulnerability aspect of quantum classifiers with comprehensive numerical simulations, which will provide valuable guidance for practical applications of using quantum classifiers to solve intricate problems where adversarial considerations are inevitable.

\section{Classical Adversarial learning and quantum classifiers: concepts and notations}\label{CALQC}

Modern technologies based on machine learning (especially deep learning) and data-driven artificial intelligence have achieved remarkable success in a broad spectrum of application domains \cite{Jordan2015Machine, Lecun2015Deep},  ranging from face/speech recognition, spam/malware detection, language translation, to self-driving cars and  autonomous robots, etc. This success raises the illusion 
that machine learning is currently at a state to be applied robustly and reliably on virtually any tasks. Yet,
as machine learning has found its way from labs to real world, the security and integrity of its applications leads to more and more serious concerns as well, especially for these applications in safety and security-critical environments \cite{Huang2011Adversarial,biggio2018wild,miller2019adversarial}, such as self-driving cars, malware detection, biometric authentication and  medical diagnostics \cite{finlayson2019adversarial}. For instance, the sign recognition system of a self-driving car may misclassify a stop sign with a little dirt on it as a parking prohibition sign, and subsequently result in a catastrophic accident. In medical diagnostics, a deep neural network may incorrectly identify a slightly-modified dermatoscopic image of a benign melanocytic nevus as malignant with even $100\%$ confidence \cite{finlayson2018adversarial}, leading to a possible medical disaster. To address these crucial concerns and problems, a new field of adversarial machine learning has emerged to study vulnerabilities of different machine learning approaches in various adversarial settings and to develop appropriate techniques to make learning more robust to adversarial manipulations \cite{vorobeychik2018adversarial}. 

This field has attracted considerable attention and is growing rapidly. In this paper, we take one step further to study the vulnerabilities of quantum classifiers and possible strategies to make them more robust to adversarial perturbations. 
For simplicity and concreteness, we will only focus our discussion on supervised learning scenarios, although a generalization to unsupervised cases is possible and worth systematic future investigations.  We start with a brief introduction to the basic concepts, notations, and ideas of classical adversarial learning and quantum classifiers. 
In supervised learning, the training data is labeled beforehand:
\(\mathcal{D}_N=\{(\mathbf{x}^{(1)},y^{(1)}),\cdots,(\mathbf{x}^{(N)},y^{(N)})\}\),  where \(\mathbf{x}^{(i)}\) ($i=1,\cdots,N$) is the data to be classified and \(y^{(i)}\) denotes its corresponding label. 
The essential task of supervised learning is to learn from the labeled data a model $y=h(\mathbf{x};\eta)$ (a classifier) that provides a general rule on how to assign labels to data outside the training set \cite{goodfellow2016deep}. This is usually accomplished by minimizing certain loss function over some set of model parameters that are collectively denoted as $\eta$: $\min_{\eta} \mathcal{L}_N(\eta)$, where  $\mathcal{L}_N(\mathbf{\eta})=\frac{1}{N}\sum_{i=1}^N L(h(\mathbf{x}^{(i)};\mathbf{\eta}),y^{(i)})$ denotes the averaged loss function over the training data set. To solve this minimization problem, different loss functions and optimization methods have been developed. Each of them bearing its own advantages and disadvantages, and the choice of which one to use depends on the specific problem. 

Unlike training the classifiers, generating adversarial examples is a different process, where we consider the model parameters $\eta$ as fixed and instead  optimize over the input space. More specifically, we search for a perturbation $\delta$ within a small region $\Delta$, which can be added into the input sample $\mathbf{x}^{(i)}$ so as to {\it maximize} the loss function:
\begin{equation}
\max_{\delta\in\Delta} \;L(h(\mathbf{x}^{(i)}+\delta;\eta),y^{(i)}),\label{eq:AdvLmaxLoss}
\end{equation}
Here in order to ensure that the adversarial perturbation is not completely changing the input data, we constrain $\delta$ to be from a small region $\Delta$, the choice of which is domain-specific and vitally depends on the problem under consideration. 
A widely adopted choice of  $\Delta$ is the $\ell_p$-norm bound: $||\delta||_p\leq \epsilon$, where the \(\ell_p\)-norm is defined as: \(\|x\|_{p}=\left(\sum_{i=1}^{N}\left\|x_{i}\right\|^{p}\right)^{\frac{1}{p}}.\) In addition, since there is more than one way to attack machine learning systems, different classification schemes of the attacking strategies have been proposed in adversarial machine learning \cite{vorobeychik2018adversarial,li2018security,miller2019adversarial,chakraborty2018adversarial}. Here, we follow Ref.~\cite{vorobeychik2018adversarial} and classify attacks along the following three dimensions: timing (considering when the attack takes place, such as attacks on models vs. on algorithms), information (considering what information the attacker has about the learning model or algorithm, such as white-box vs. black-box attacks), and goals (considering different reasons for attacking, such as targeted vs. untargeted attacks).  We will not attempt to exhaust all possible attacking scenarios, which is implausible due to its vastness and complexity. Instead, we only focus on several types of attacks that have already capture the essential messages we want to deliver in this paper. In particular, along the ``information" dimension, we  consider white-box and black-box attacks. In the white-box setting, the attacker has full information about the learned model and the learning algorithm, whereas the black-box setting assumes that the adversary does not have precise information about either the model or the algorithm used by the learner. In general, obtaining adversarial examples in the black-box setting is more challenging. Along the ``goals" dimension, we distinguish two major categories: targeted and untargeted attacks.  
In a targeted attack, the attacker aims to deceive the classifier into outputting a particularly targeted label. In contrast, untargeted attacks (also called reliability attacks in the literature) just attempt to cause the classifier make erroneous predictions, but no particular class is aimed. We also mention that a number of different methods have been proposed to solve the optimization problem in Eq.~\eqref{eq:AdvLmaxLoss} or its variants in different scenarios \cite{biggio2018wild}. We refer to Refs. \cite{biggio2018wild,madry2017towards,CW,BIM,papernot2015limitations,
vorobeychik2018adversarial,szegedy2013intriguing,MIM,goodfellow2014explaining,papernot2016transferability, papernot2017practical,chen2017zoo} for more technique details. 
As for our purpose, we will mainly explore  the fast gradient sign method (FGSM) \cite{madry2017towards}, basic iterative method (BIM)\cite{BIM}, projected gradient descent (PGD) \cite{madry2017towards}, and momentum iterative method (MIM)\cite{MIM} in the white-box setting and the transfer attack \cite{goodfellow2014explaining}, substitute model attack \cite{papernot2016transferability, papernot2017practical}, and zeroth-order optimization (ZOO) attack \cite{chen2017zoo} methods in the black-box setting.

On the other hand, another major motivation for studying adversarial learning is to develop proper defense strategies to enhance the robustness of machine learning systems to adversarial attacks. 
Along this direction, a number of countermeasures have been proposed as well in recent years \cite{vorobeychik2018adversarial}.   For instance, Kurakin {\it et al.} introduced the idea of adversarial training \cite{kurakin2016adversarial}, where the robustness of the targeted classifier is enhanced by retraining with both the original legitimate data and the crafted data.  Samangouei {\it et al.} came up with a mechanism \cite{samangouei2018defense} that uses generative adversarial network \cite{Goodfellow2014Generativev} as a countermeasure for adversarial perturbations. Papernot {\it et al.} proposed a defensive mechanism \cite{papernot2016distillation} against adversarial examples based on distilling knowledge in neural networks \cite{hinton2015distilling}. Each of these proposed defense mechanisms works notably well against particular classes of attacks, but none of them could be used as a generic solution for all kinds of attacks.  In fact, we {\it cannot} expect a universal defense strategy that can make all machine learning systems robust to all types of attacks, as one strategy that closes a certain kind of attack will unavoidably open another vulnerability for other types of attacks which exploit the underlying defense mechanism. In this work, we will use adversarial learning to enhance the robustness of quantum classifiers against certain types of adversarial perturbations.

Quantum classifiers are counterparts of classical ones that run on quantum devices. In recent years, a number of different approaches have been proposed to construct efficient quantum classifiers \cite{schuld2018circuit,farhi2018classification,schuld2017implementing,mitarai2018quantum,schuld2019quantum,havlivcek2019supervised,zhu2018training,cong2018quantum,wan2017quantum,grant2018hierarchical,du2018implementable,uvarov2019machine,blank2019quantum,rebentrost2014quantum,tacchino2019artificial,uvarov2019machine}, with some of them even been implemented in proof-of-principle experiments. One straightforward construction, called the quantum variational classifier \cite{farhi2018classification,mitarai2018quantum,schuld2018circuit}, is to use a variational quantum circuit to classify the data in a way analogous to the classical support vector machines \cite{goodfellow2016deep}.  Variants of this type of classifiers include hierarchical quantum classifiers \cite{grant2018hierarchical} (such as these inspired by the structure of tree tensor network or  multi-scale entanglement renormalization ansatz) and quantum convolutional neural networks \cite{cong2018quantum}. Another approach, called the quantum kernel \cite{schuld2019quantum,havlivcek2019supervised,blank2019quantum}, utilizes the quantum Hilbert space as the feature space for data and compute the kernel function via quantum devices. Both the quantum variational classifier and the quantum kernel approach have been demonstrated in a recent experiment with superconducting qubits \cite{havlivcek2019supervised}. In addition, hierarchical quantum classifiers have also been implemented by using the IBM quantum experience \cite{IBMQuantumExperience} and their robustness to depolarizing noises has been demonstrated in principle \cite{grant2018hierarchical}. These experiments showcase the intriguing potentials of using the noisy intermediate-scale quantum devices \cite{Preskill2018quantumcomputingin} (which are widely expected to be available in the near future) to solve practical machine learning problems, although an unambiguous demonstration of quantum advantages is still lacking.  Despite these exciting progresses, an important question of both theoretical and experimental relevance concerning the reliability of quantum classifiers remains largely unexplored: are they robust to adversarial perturbations?

\section{Vulnerability of quantum classifiers}\label{VQC}

As advertised in the above discussion, quantum classifiers are vulnerable to adversarial perturbations. In this section, we will first introduce the general structure of the quantum classifiers and the learning algorithms used in this paper and several attacking methods to obtain adversarial perturbations with technique details provided in the Appendix. We then apply these methods to concrete examples to explicitly show the vulnerability of quantum classifiers in diverse scenarios, including quantum adversarial learning of real-life images (e.g., handwritten digit images in MNIST), topological phases of matter, and quantum data from the ground states of physical Hamiltonians.

\begin{figure}
\centering
\includegraphics[width=0.48\textwidth]{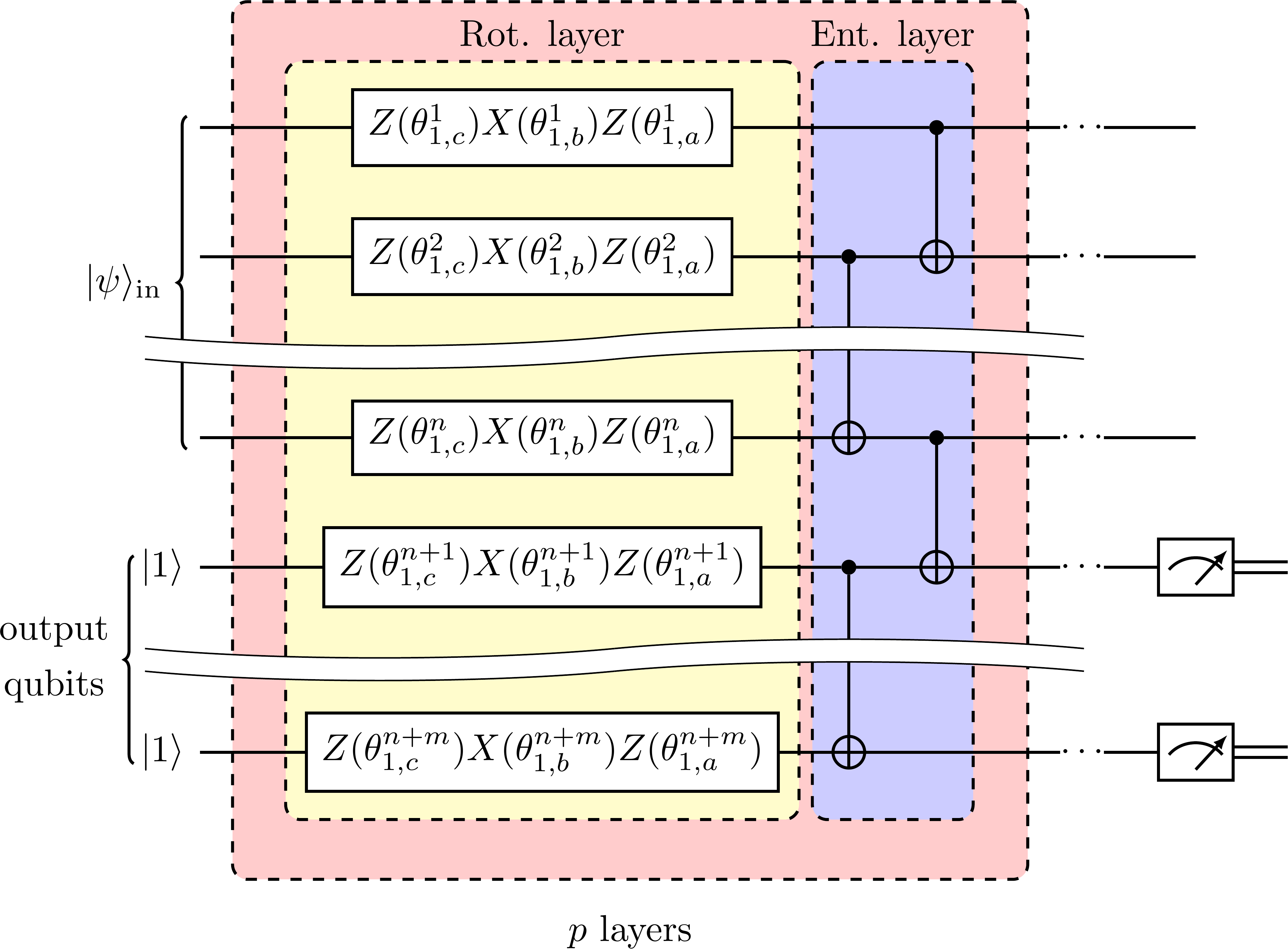}
\caption{\label{fig:circuitclassifier} The sketch of a quantum circuit classifier. The classifier consists of $p$ layers, with each layer containing a rotation unit and an entangler unit.  The rotation unit performs arbitrary single-qubit Euler rotations implemented as a combination of $Z$ and $X$ gates: $U_{q,i}({\bm{\theta}})=Z_{\theta_{q,i}^c}X_{\theta_{q,i}^b}Z_{\theta_{q,i}^a}$ with $\mathbf{\theta}$ representing the Euler angles, $q$ identifying the qubit, and $i=1,2,\cdots, p$ referring to the label of layers. The entangler unit entangles all qubits and is composed of a series of CNOT gates. The initial state $|\psi\rangle_{\text{in}}$, which is a $n$-qubit state, encodes the complete information of the input data to be classified. The projection measurement on the output qubits give the predicting probability for each category and the input data is assigned a label that bearing the largest probability.
}\label{CircuitQC}
\end{figure}

\subsection{Quantum classifiers: training and adversarial attacks}\label{quantum-classifier-training-and-adversarial-attack}

Quantum classifiers take quantum states as input. Thus, when they are used to classify classical data, we need first to convert classical data into quantum states. This can be done with an encoding operation, which basically implements a feature map from the $D$-dimensional Euclidean space (where the class data is typically represented by $D$-dimensional vectors) to the $2^n$-dimensional Hilbert space for $n$ qubits: $\varphi: \mathbb{R}^D\to \mathbb{C}^{2^n}$.  There are two common ways of encoding classical data into quantum states: amplitude encoding and qubit encoding  \cite{rebentrost2014quantum, kerenidis2016quantum, schuld2018circuit, schuld2017implementing, giovannetti2008architectures, Harrow2009Quantum, lloyd2013quantum, wiebe2014quantum, cong2016quantum, giovannetti2008quantum, rebentrost2014quantum, Aaronson2015Read}. Amplitude encoder maps an input vector \(\mathbf{x}\in \mathbb{R}^D\) (with some possible preprocessing such as normalization) directly into the amplitudes of the $2^n$-dimensional ket vector $|\psi\rangle_{\text{in}}$ for $n$ qubits in the computational basis. Here, for simplicity, we assume that \(D\) is a power of two such that we can use \(D=2^n\) amplitudes of a \(n\)-qubit system (in fact, if $D<2^n$ we can add $2^n-D$ zeros at the end of the input vector to make it of length $2^n$).  Such a converting procedure can be achieved with a circuit whose depth is linear in the number of features in the input vectors with the routines in Refs. \cite{mottonen2004quantum, knill1995approximation, plesch2011quantum}. With certain approximation or structure, the required overhead can be reduced to polylogarithmic in $D$ \cite{grover2002creating, soklakov2006efficient}. This encoding operation can also be made more efficient by using more complicated approaches such as tensorial feature maps \cite{schuld2018circuit}.
Qubit encoder, in contrast, uses $D$ (rather than $O(\log D$) as in amplitude encoding) qubits to encode the input vector. We first rescale the data vectors element-wise to lie in \([0,\frac{\pi}{2}]\) and encode each element with a qubit using the following scheme: \(|\phi_d\rangle = \cos(x_d)|0\rangle+\sin(x_d)|1\rangle\), where $x_d$ is the $d$-th element of the rescaled vector. The total quantum input state that encodes the data vectors is then a tensor product $|\phi\rangle=\otimes_{d=1}^D |\phi_d\rangle$.  Qubit encoding does not require a quantum random access memory \cite{giovannetti2008quantum}  or a complicated circuit to prepare the highly entangled state $|\psi\rangle_{\text{in}}$, but it demands much more qubits to perform the encoding and hence is more challenging to numerically simulate the training and adversarial attacking processes on a classical computer. As a result, we will only focus on amplitude encoding in this work, but the generalization to other encoding schemes is straightforward and worth investigation in the future.

We choose a hardware-efficient quantum circuit classification model, which has been used as a variational quantum eigensolver for small molecules and quantum magnets in a recent experiment with superconducting qubits \cite{kandala2017hardware}. The schematic illustration of the model is shown in Fig. \ref{fig:circuitclassifier}. Without loss of generality, we assume that the number of categories to be classified is $K$ and each class is labeled by an integer number $1\leq k\leq K$. We use $m$  qubits ($2^{m-1}<K\leq 2^m$) to serve as output qubits that encode the category labels. A convenient encoding strategy that turns discrete labels into a vector of real numbers is the so-called one-hot encoding \cite{goodfellow2016deep}, which converts a discrete input value $0<k\leq K$ into a vector $\mathbf{a}\equiv (a_1,\cdots,a_{2^m})$ of  length $2^m$ with $a_k=1$ and $a_j=0$ for $j\neq k$. For the convenience of presentation, we will use $y$ and $\mathbf{a}$ interchangeably to denote the labels throughout the rest of the paper.    In such a circuit model , we first prepare the input state to be \(|\psi\rangle_{\text{in}}\otimes |1\rangle^{\otimes m}\) with $|\psi\rangle_{\text{in}}$ a $n$-qubit state encoding the complete information of the data to be classified,  and then apply a unitary transform consisting of \(p\) layers of interleaved operations. Each layer contains a rotation unit that performs arbitrary single-qubit Euler rotations and an entangler layer that generates entanglement between qubits. This generates  a variational wavefunction \(|\Psi({\Theta})\rangle=\prod_{i=1}^pU_i(|\psi\rangle_{\text{in}}\otimes |1\rangle^{\otimes m})\), where $U_i=[\prod_q U^q_{i}({\bm{\theta}}_i)] U_{\text{ENT}}=(\prod_q Z_{\theta^{q}_{i,c}}X_{\theta^{q}_{i,b}}Z_{\theta^{q}_{i,a}}) U_{\text{ENT}}$ denotes the unitary operation for the $i$-th layer. Here $U_{\text{ENT}}$ represents the unitary operation generated by the entangler unit and we use $\bm{\theta}_i$ to denote collectively all the parameters in the $i$-th layer and $\Theta$ to denote collectively all the parameters evolved in the whole model.  We mention that the arbitrary single-qubit rotation together with the control-NOT gate gives a universal gate set in quantum computation. Hence our choice of this circuit classifier is  universal as well, in the sense that it can approximate any desired function as long as $p$ is large enough. One may choose other models, such as hierarchical quantum classifiers \cite{grant2018hierarchical} or the quantum convolutional neural network \cite{cong2018quantum}, and we expect that the attacking methods and the general conclusion should carry over straightforwardly to these models.

During the training process, the variational parameters $\Theta$ will be updated iteratively so as to minimize certain loss functions. The measurement statistics on the output qubits will determine the predicted label for the input data encoded in state \(|\psi\rangle_{\text{in}}\). For example, in the case of two-category classification, we can use $y\in\{0,1\}$ to label the two categories and  the number of output qubits is one.   We estimate the probability for each class by measuring the expectation values of the projections: $P(y=l)=\text{Tr}(\rho_{\text{out}}|l\rangle\langle l|)$, where $l=0,1$ and $\rho_{\text{out}}=\text{Tr}_{1,\cdots,n}(|\Psi(\Theta)\rangle\langle\Psi(\Theta)|)$ is the reduced density matrix for the output qubit. We assign a label $y=0$ to the data sample $\mathbf{x}$ if $P(y=0)$ is larger than $P(y=1)$ and say that $\mathbf{x}$ is classified to be in the $0$ category with probability $P(y=0)$ by the classifier. The generalization to multi-category classification is straightforward. One observation which may simplify the numerical simulations a bit is that the diagonal elements of  $\rho_{\text{out}}$, denoted as $\mathbf{g}\equiv (g_1,\cdots,g_{2^m})=\text{diag}(\rho_{\text{out}})$, in fact give all the probabilities for the corresponding categories. 
 
In classical machine learning, a number of different loss functions have been introduced for training the networks and characterizing their performances. Different loss functions possess their own pros and cons and are best suitable for different problems. For our purpose, we define the following loss function based on cross-entropy for a single data sample encoded as $|\psi\rangle_{\text{in}}$:
\begin{equation}
L(h(|\psi\rangle_{\text{in}};\Theta),\mathbf{a})=-\sum_k a_k \log g_k.
\label{eq:loss}
\end{equation}
During the training process, a classical optimizer is used to search for the optimal parameters $\Theta^*$ that minimize the averaged loss function over the training data set: $\mathcal{L}_N(\Theta)=\frac{1}{N}\sum_{i=1}^N L(h(|\psi\rangle_{\text{in}}^{(i)};\Theta),\mathbf{a}^{(i)})$. Various gradient descent algorithms, such as the stochastic gradient descent \cite{sweke2019stochastic} and   quantum natural gradient descent \cite{yamamoto2019natural, stokes2019quantum}, etc., can be employed to do the optimization. We use Adam \cite{kingma2014adam, reddi2018convergence}, which is an adaptive learning rate optimization algorithm  designed specifically for training deep neural networks, to train the quantum classifiers.  

A crucial quantity that plays a vital role in minimizing  $\mathcal{L}_N(\Theta)$ is its gradient with respect to model parameters. Interestingly, owing to the special structures of our quantum classifiers this quantity can be directly obtained from the projection measurements through the following equality \cite{liu2018differentiable}: 
\begin{equation}
\frac{\partial\langle \mathcal{L}_N(\Theta)\rangle_{\vartheta}}{\partial \vartheta}=\frac{1}{2}\left(\langle \mathcal{L}_N(\Theta)\rangle_{\vartheta+\frac{\pi}{2}}-\langle \mathcal{L}_N(\Theta)\rangle_{\vartheta-\frac{\pi}{2}}\right),
\label{eq:gradient}
\end{equation}
where $\vartheta$ denotes an arbitrary single parameter in our circuit classifier and $\langle \mathcal{L}_N(\Theta) \rangle_{\xi} $ ($\xi=\vartheta, \vartheta+\frac{\pi}{2}, \text{ and } \vartheta-\frac{\pi}{2}$) represents the expectation value of $\mathcal{L}_N(\Theta)$ with the corresponding parameter set to be $\xi$. We note that the equality in Eq. \eqref{eq:gradient} is exact, in sharp contrast to other models for quantum variational classifiers where the gradients can only be approximated by finite-difference methods in general. It has been proved that an accurate gradient based on quantum measurements could lead to substantially faster convergence to the optimum in many scenarios \cite{harrow2019low}, in comparison with the finite-difference method approach.

\begin{figure}
\centering
\includegraphics[width=0.48\textwidth]{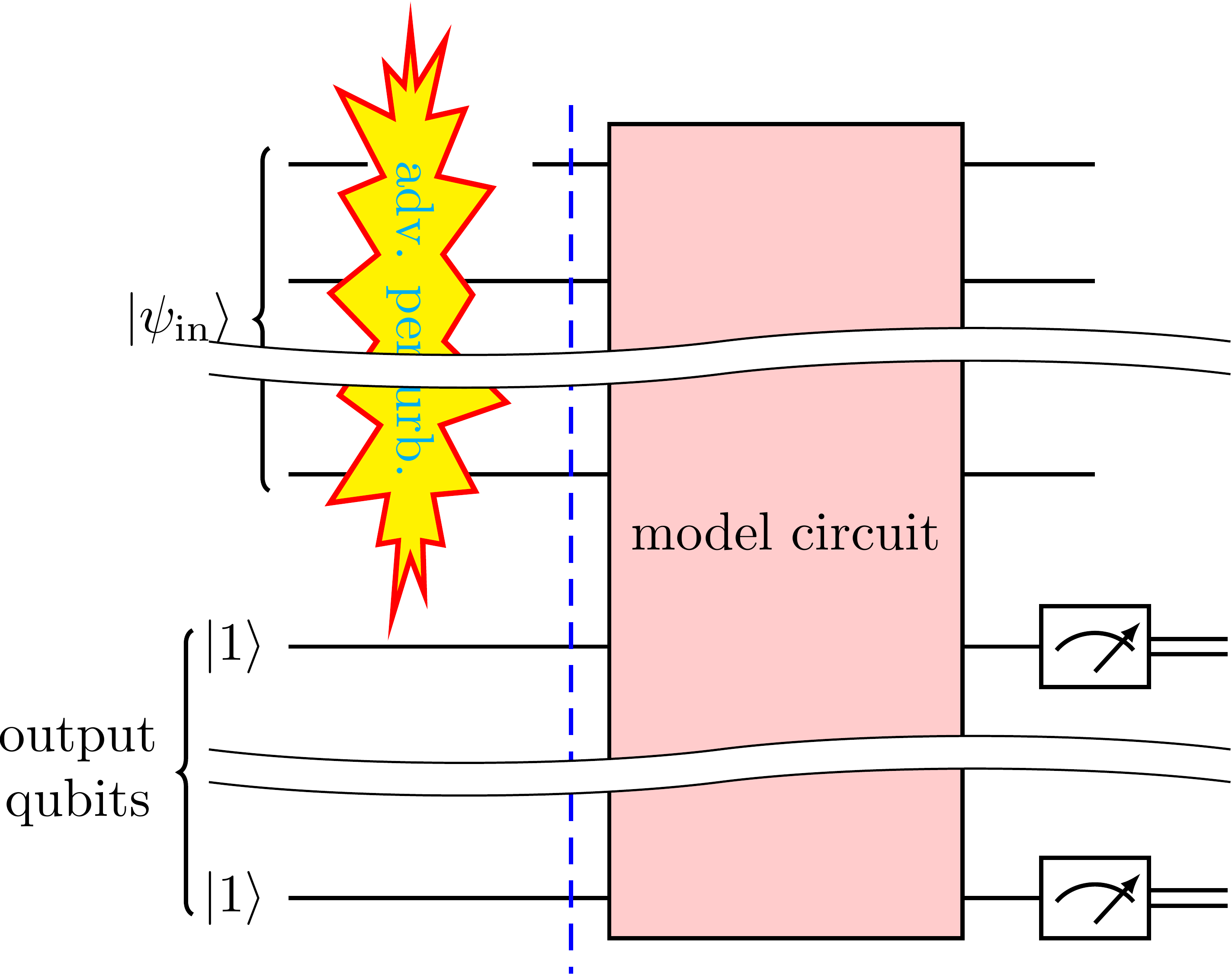}
\caption{\label{fig:adv-circuit} A sketch of adding adversarial perturbations to the input data for quantum classifiers. Throughout this paper, we mainly focus on evasion attack \cite{vorobeychik2018adversarial}, which is the most common type of attack in adversarial learning.  In this setting, the attacker attempts to deceive the quantum classifiers by adjusting malicious samples during the testing phase. Adding a tiny amount of adversarial noise can cause quantum classifiers to make incorrect predictions.}
\end{figure}

We now give a general recipe on how to generate adversarial perturbations for quantum classifiers. Similar to the case of classical adversarial learning, this task essentially reduces to another optimization problem where we search for a small perturbation within an appropriate region $\Delta$ that can be added into the input data so that the loss function is maximized. A quantum classifier can classify both classical and quantum data. Yet, adding perturbations to classical data is equivalent to adding perturbations to the initial quantum state $|\psi\rangle_{\text{in}}$. Hence, it is sufficient to consider only perturbations to $|\psi\rangle_{\text{in}}$, regardless of whether the data to be classified is quantum or classical. A pictorial illustration of adding adversarial perturbations to the input data for a quantum classifier is shown in Fig. \ref{fig:adv-circuit}. In the case of untargeted attacks, we attempt to search a perturbation operator $U_{\delta}$ acting on $|\psi\rangle_{\text{in}}$ to maximize the loss function:
\begin{equation}
U_{\delta}\equiv \argmax_{U_{\delta}\in \Delta} L(h( U_{\delta}|\psi\rangle_{\text{in}};\Theta^*),\mathbf{a}),\label{Eq:WBunT}
\end{equation}
where $\Theta^*$ denotes the fixed parameters determined during the training process, $|\psi\rangle_{\text{in}}$ encodes the information of the data sample $\mathbf{x}$ supposed to be under attack, and $\mathbf{a}$ represents the correct label for $\mathbf{x}$ in the form of one-hot encoding. On the other hand, in the case of targeted attacks we aim to search a perturbation $U_{\delta}^{(\text{t})}$ that minimizes (rather than maximizes) the loss function under the condition that the predicted label is targeted to be a particular one:
\begin{eqnarray}
U_{\delta}^{(\text{t})}\equiv \argmin_{U_{\delta}^{(\text{t})}\in \Delta} L(h( U_{\delta}^{(\text{t})}|\psi\rangle_{\text{in}};\Theta^*),\mathbf{a}^{(\text{t})}),
\label{eq:adversarial-learning-targeted}
\end{eqnarray}
where $\mathbf{a}^{(\text{t})}$ is the targeted label that is different from the correct one $\mathbf{a}\neq \mathbf{a}^{(\text{t})}$.

In general, $\Delta$ can be a set of all unitaries that are close to the identity operator. This corresponds to the additive attack in classical adversarial machine learning, where we modify each component of the data vector independently. In our simulations, we use automatic differentiation \cite{rall1996introduction}, which computes derivatives to machine precision,  to implement this type of attack. In addition,   for simplicity we can further restrict $\Delta$ to be a set of products of local unitaries that are close to the identity operator. This corresponds to the functional adversarial attack \cite{mcclean2017hybrid} in classical machine learning. It is clear that the searching space for the functional attack  is much smaller than that for the additive attack and one may regard the former as a special case for the later.

We numerically simulate the training and inference process of the quantum classifiers on a classical cluster by using the Julia language \cite{bezanson2017julia} and  Yao.jl  \cite{Yao} framework. We run the simulation parallelly on the CPUs or GPUs, depending on different scenarios. The parallel nature of the mini-batch gradient descent algorithm naturally fits the merits of GPUs and thus we use CuYao.jl \cite{CuYao}, which is a very efficient GPU implementation of Yao.jl \cite{Yao}, to gain speedups for the cases that are more resource-consuming.    We find that the performance of calculating mini-batch gradients on a single GPU is ten times better than that of parallelly running on CPUs with forty cores. The automatic differentiation is implemented with  Flux.jl \cite{Innes2018Flux} and Zygote.jl \cite{Zygote}. Based on this implementation, we can optimize over a large number of parameters for circuit depth as large as \(p=50\). In general, we find that increases in circuit depth (model capacity) are conducive to the achieved accuracy. We check that the model does not overfit because the loss of the training data set and validation data set is close. So there is no need for introducing regularization techniques such as Dropout \cite{srivastava2014dropout} to avoid overfitting.

Now we have introduced the general structure of our quantum classifiers and the methods to train them and to obtain adversarial perturbations. In the following subsections, we will demonstrate how these methods work by giving three concrete examples. These examples explicitly showcase the extreme vulnerability of quantum classifiers.

\begin{figure}
\centering
\includegraphics[width=0.99\linewidth]{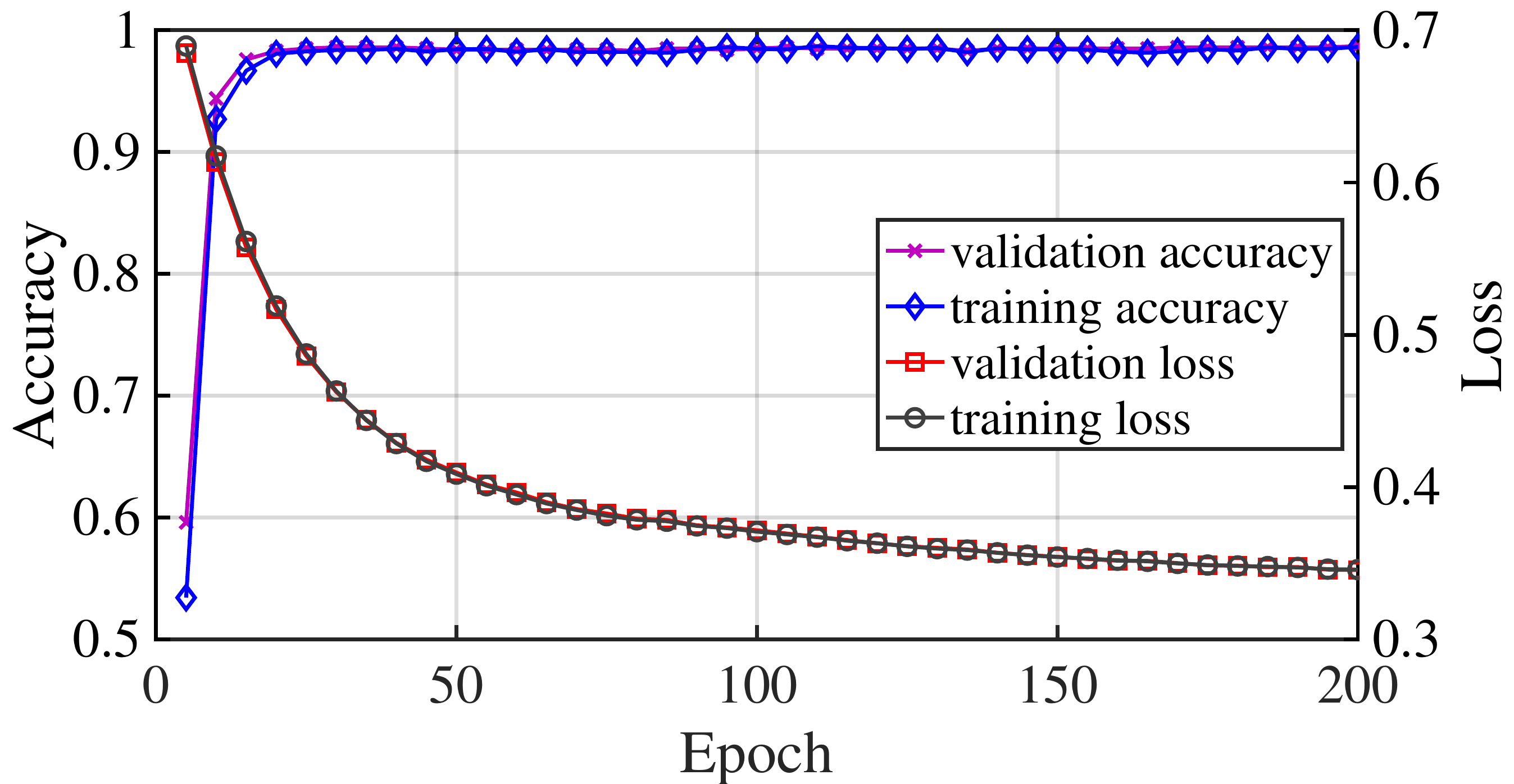}
\caption{\label{fig:acc-loss-two} The average accuracy and loss as a function of the number of training steps. We use a depth-10 quantum classifier with structures shown in Fig. \ref{CircuitQC} to perform binary classification for images of digits $1$ and $9$ in MNIST. To train the classifier, we use the Adam optimizer with a batch size of $256$ and a learning rate of $0.005$ to minimize the loss function in Eq. \eqref{eq:loss}. The accuracy and loss are averaged on 11633 training samples and 1058 validation samples (which are not contained in the training dataset).  }
\end{figure}

\subsection{Quantum adversarial learning images}\label{quantum-adversarial-learning-images}

Quantum information processors possess unique properties such as quantum parallelism and quantum superposition, making them intriguing candidates for speeding up image recognitions in machine learning. 
It has been shown that some quantum image processing algorithms may achieve exponential speedups over their classical counterparts \cite{venegas2003storing, yao2017quantum}. Researchers have employed quantum classifiers for many different image data sets \cite{schuld2018circuit}. Here, we focus on the MNIST handwritten digit classification dataset~\cite{mnist}, which is widely considered to be a real-life testbed for new machine learning paradigms. For this dataset, near-perfect results have been reached using various classical supervised learning algorithms~\cite{arewethereyet}. The MNIST data set consists of hand-drawn digits, from 0 through 9 in the form of gray-scale images. Each image is two dimensional, and contains \(28 \times 28\) pixels. Each pixel of an image in the dataset has a pixel-value, which is an integer ranging from $0$ to $255$ with $0$ meaning the darkest and $255$ the whitest color.  For our purpose, we slightly reduced the size of the images from \(28\times28\) pixels to \(16\times 16\) pixels, so that we can simulate the training and attacking processes of the quantum classifier with moderate classical  computational  resources. In addition, we normalize these pixel values and encode them into a pure quantum state using the amplitude encoding method mentioned in Sec. \ref{quantum-classifier-training-and-adversarial-attack}.

We first train the quantum classifiers to identify different images in the MNIST with sufficient classification accuracy. The first case we consider is a two-category classification problem, where we aim to classify the images of digits $1$ and $9$ by a quantum classifier with structures introduced shown in Fig. \ref{CircuitQC} . From the MNIST dataset, we select out all images of $1$ and $9$ to form a sub-dataset, which contains a training dataset of size 11633 (used for training the quantum classifier), a validation dataset of size $1058$ (used for tuning hyperparameters, such as the learning rate), and a testing set of size $2144$ (used for evaluating the final performance of the quantum classifier). In Fig. ~\ref{fig:acc-loss-two}, we plot the average accuracy and loss for the training and validation datasets respectively as a function of the number of epochs. From this figure, the accuracy for both the training and validation increases rapidly at the beginning of the training process and then saturate at a high value of $\approx 98\%$. Meanwhile, the average loss for both training and validation decreases as the number of epochs increases. The difference between the training loss and validation loss is very small, indicating that the model does not overfit. In addition,  the performance of the quantum classifier is also tested on the testing set and we find that our classifier can achieve a notable accuracy of $98\%$ after around fifteen epochs.

\begin{figure}
\centering
\includegraphics[width=0.48\textwidth]{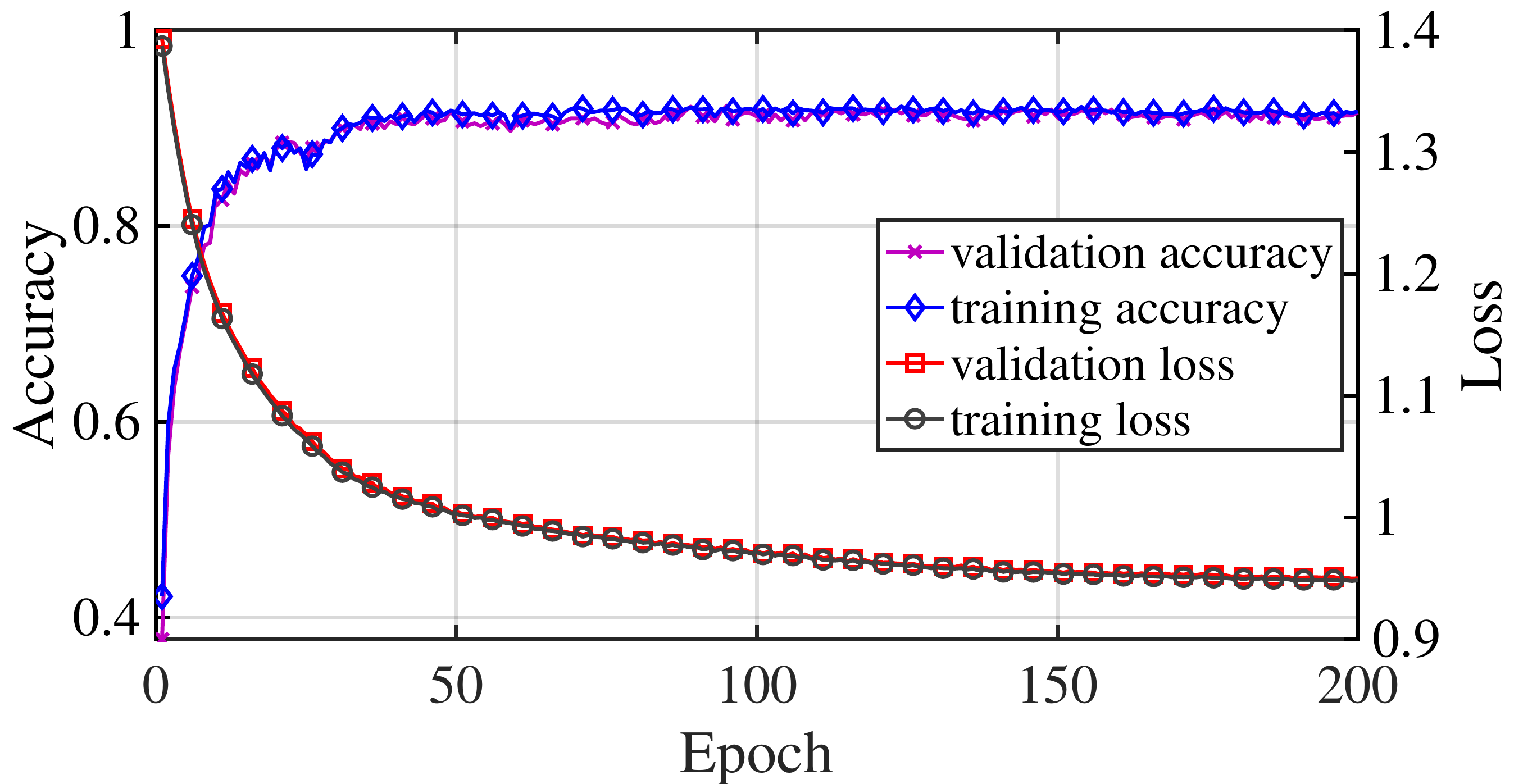}
\caption{\label{fig:acc-loss-four}The average accuracy and loss for the four-category quantum classifier as a function of the number of epochs. Here, we use a quantum classifier with structures shown in Fig.~\ref{CircuitQC} and depth forty ($p=40$) to  perform multi-class classification for images of digits $1$, $3$, $7$, and $9$. To train the classifier, we use the Adam optimizer with a batch size of $512$ and learning rate of $0.005$ to minimize the loss function in Eq.~\eqref{eq:loss}. The accuracy and loss are averaged on $20000$ training samples and $2000$ validation samples. }
\end{figure}

For two-category classifications, the distinction between targeted and untargeted attacks blurs since the target label can only be simply the alternative label. Hence, in order to illustrate the vulnerability of quantum classifiers under targeted attacks, we also need to consider a case of multi-category classification. To this end, we train a quantum classifier to distinguish four categories of handwritten digits: $1$, $3$, $7$, and $9$. Our results are plotted Fig. ~\ref{fig:acc-loss-four}. Similar to the case of two-category classification, we find that both the training and validation accuracies increase rapidly at the beginning of the training process and then saturate at a value of $\approx 92\%$, which is smaller than that for the two-category case. After training, the classifier is capable of predicting the corresponding digits for the testing dataset with an accuracy of $91.6\%$. We mention that one can further increase the accuracy for both the two- and four-category classifications, by using the original $28\times 28$-pixel images in MNIST or using a quantum classifier with more layers. But this demands more computational resources.

After training, we now fix the parameters of the corresponding quantum classifiers and study the problem of how to generate adversarial examples in different situations. We consider both the white-box and black-box attack scenarios. For the white-box scenario, we explore both untargeted and targeted attacks. For the black-box scenario, we first generate adversarial examples for classical classifiers and show that quantum classifiers are also vulnerable to these examples owing to the transferability properties of adversarial examples.

\begin{figure}
\centering
\includegraphics[width=0.48\textwidth]{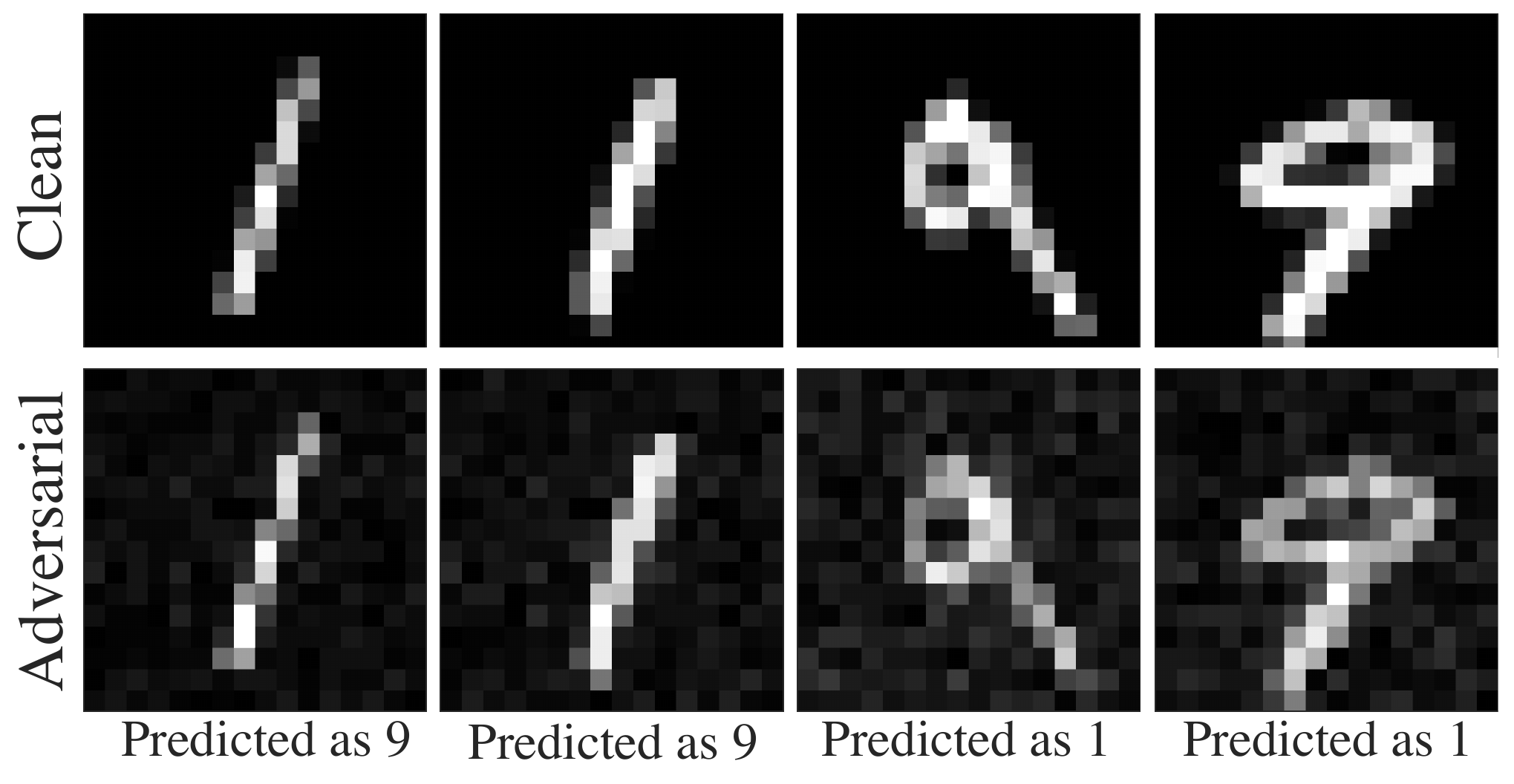}
\caption{\label{fig:adversarial-examples-1-9} The clean and the corresponding adversarial images for the quantum classifier generated by the basic iterative method (see Appendix). Here, we apply the additive attack in the white-box untargeted setting. For the legitimate clean images, the quantum classifier can correctly predict their labels with confidence larger than $78\%$. After attacks, the classifier will misclassify the crafted images of digit $1$ ($9$) as digit $9$ ($1$) with notably  high confidence, although the differences between the crafted and clean images are almost imperceptible to human eyes. In fact, the average fidelity is $0.916$, which is very close to unity.}
\end{figure}

\subsubsection{White-box attack: untargeted}\label{white-box-attack-untargeted}

In the white-box setting, the attacker has full information about the quantum classifiers and the learning algorithms. In particular, the attacker knows the loss function that has been used and hence can calculate its gradients with respect to the parameters that characterize the perturbations. As a consequence, we can use different gradient-based methods developed in the classical adversarial machine learning literature, such as the FGSM~\cite{madry2017towards}, BIM~\cite{BIM}, PGD~\cite{madry2017towards}, and MIM~\cite{MIM}, to generate adversarial examples. For untargeted attacks, the attacker only attempts to cause the classifier to make incorrect predictions, but no particular class is aimed. In classical adversarial learning, a well-known example in the white-box untargeted scenario concerns facial biometric systems \cite{sharif2016accessorize}, whereby wearing a pair of carefully-crafted eyeglasses  the attacker can have her face misidentified by the state-of-the-art face-recognition system as any other arbitrary face (dodging attacks). Here, we show that quantum classifiers are vulnerable to such attacks as well.

\begin{figure}[htbp!]
\centering
\includegraphics[width=0.48\textwidth]{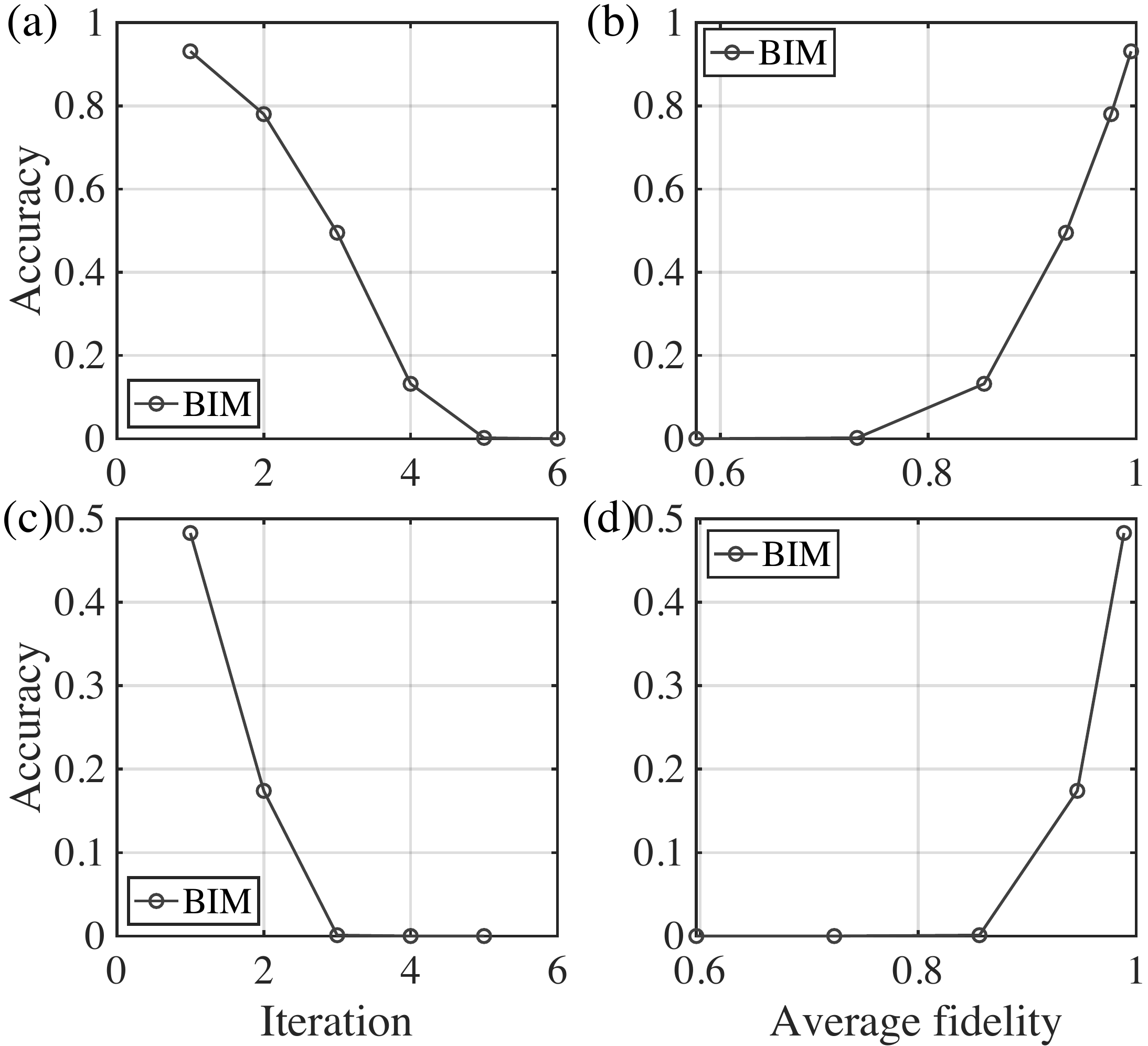}
\caption{\label{fig:M2_M_4_acc_fid} Effect of adversarial untargeted additive attacks on the accuracy of the quantum classifier for the problem of classifying handwritten digits. We use the basic iterative method to obtain adversarial examples. The circuit depth of the model is 20. We choose the step size as 0.1. (a)-(b) For the classifier that classifies digit 1 and 9, accuracy decreases as the average fidelity between the adversarial samples and clean samples decreases. Accuracy decreases as we increase the number of iterations of the attacking algorithm. (c)-(d) Similar plots for the problem of classifying four digits 1, 3, 7, and 9. }
\end{figure}

For the simplest illustration, we first consider attacking additively the two-category quantum classifier discussed above in the withe-box untargeted setting. In Fig. ~\ref{fig:adversarial-examples-1-9}, we randomly choose samples for digits $1$ and $9$ from MNIST and then solve the Eq. \eqref{Eq:WBunT} iteratively by the BIM method to obtain their corresponding adversarial examples. This figure shows the original clean images and their corresponding adversarial ones for the two-category quantum classifier. For these particular clean images, the quantum classifier can correctly assign their labels with confidence larger than $78\%$. Yet, after attacks the same classifier will misclassify the crafted images of digit $1$ ($9$) as digit $9$ ($1$) with decent high confidence $73\%$. 
Strikingly, the obtained adversarial examples look the same as the original legitimate samples. They only differ by a tiny amount of noise that is almost imperceptible to human eyes. To further verify that the vulnerability of the quantum classifier is not specific to particular images, but rather generic for most of (if not all) images in the dataset, we apply the same attack to all images  of digits $1$ and $9$ in the testing set of MNIST. In Fig. \ref{fig:M2_M_4_acc_fid}(a), we plot the accuracy as a function of the number of the BIM iterations. It is clear from this figure that the accuracy decreases rapidly at the beginning of the attack, indicating that more and more adjusted images are misclassified. After five BIM iterations, the accuracy decreases to zero and all adjusted images become adversarial examples misclassified by the quantum classifier. In addition, to characterize how close a clean legitimate image is to its adversarial counterpart in the quantum framework, we define the fidelity between the quantum states that encode them: $F=|\langle \psi^{\text{adv.}}|\psi^{\text{leg.}}\rangle|^2$, where $|\psi^{\text{adv.}}\rangle$ and $|\psi^{\text{leg.}}\rangle$ denote the states that encode the legitimate and adversarial sample, respectively. In Fig. \ref{fig:M2_M_4_acc_fid}(b), we compute the average fidelity at each BIM iteration and plot the accuracy as a function of average fidelity. Since the fidelity basically measures the difference between the legitimate and adversarial images, hence it is straightforward to obtain that the accuracy will decrease as the average fidelity decreases. This is explicitly demonstrated in Fig. \ref{fig:M2_M_4_acc_fid}(b). What is more interesting is that even when the accuracy decreases to zero, namely when all the adjusted images are misclassified, the average fidelity is still larger than  $0.73$. We mention that this is a fairly high average fidelity, given that the Hilbert space dimension of the quantum classifier is already very large.

\begin{table}
	\caption{\label{table:performance-of-untargeted-attack}Average fidelity ($\bar{F}$) and accuracy (in \(\%\)) of the quantum classifier when being additively attacked by the BIM  and FGSM methods in the white-box untargeted setting.  For the two-category (four-category) classification, we use a model circuit of depth $p=10$ ($p=40$). For the BIM method, we generate adversarial examples using three iterations with a step size of $0.1$. We denote such attack as BIM($3$, $0.1$). For the FGSM method, we generate adversarial examples using a single step with a step size of $0.03$ ($0.05$) for the two-category (four-category) classifier. We denote such attacks as FGSM($1$, $0.03$) and FGSM($1$, $0.05$), respectively.}
	\begin{ruledtabular}
	\begin{tabular}{llll}
		&Attacks & $\bar{F}$ & Accuracy  \\ 
		\colrule
		\multirow{2}{2em}{two-category}&BIM (3, 0.1) & 0.923 & \(15.6\%\) \\
		&FGSM (1, 0.03) & 0.901 & \(00.0\%\) \\
		\colrule
		\multirow{2}{2em}{four-category}&BIM (3, 0.1) & 0.943 & \(23.7\%\) \\ 
		&FGSM (1, 0.05) & 0.528 & \(00.0\%\) \\ 
	\end{tabular}
	\end{ruledtabular}
\end{table}

In the above discussion, we have used Eq. \eqref{Eq:WBunT}, which is suitable for the untargeted attack, to generate adversarial examples. However, the problem we considered is a two-category classification problem and the distinction between targeted and untargeted attacks is ambiguous. A more  unambiguous approach is to consider untargeted attacks to the four-category quantum classifier. Indeed, we have carried out such attacks and our results are plotted in Fig. \ref{fig:M2_M_4_acc_fid}(c-d), which are similar to the corresponding results for the two-category scenarios. Moreover, we can also consider utilizing different optimization methods to do white-box untargeted attacking for the quantum classifiers. In Table \ref{table:performance-of-untargeted-attack}, we summarize the performance of two different methods (BIM and FGSM) in attacking both the two-category and four-category quantum classifiers. Both the BIM and FGSM methods perform noticeably well.

\begin{figure}
\centering
\includegraphics[width=0.48\textwidth]{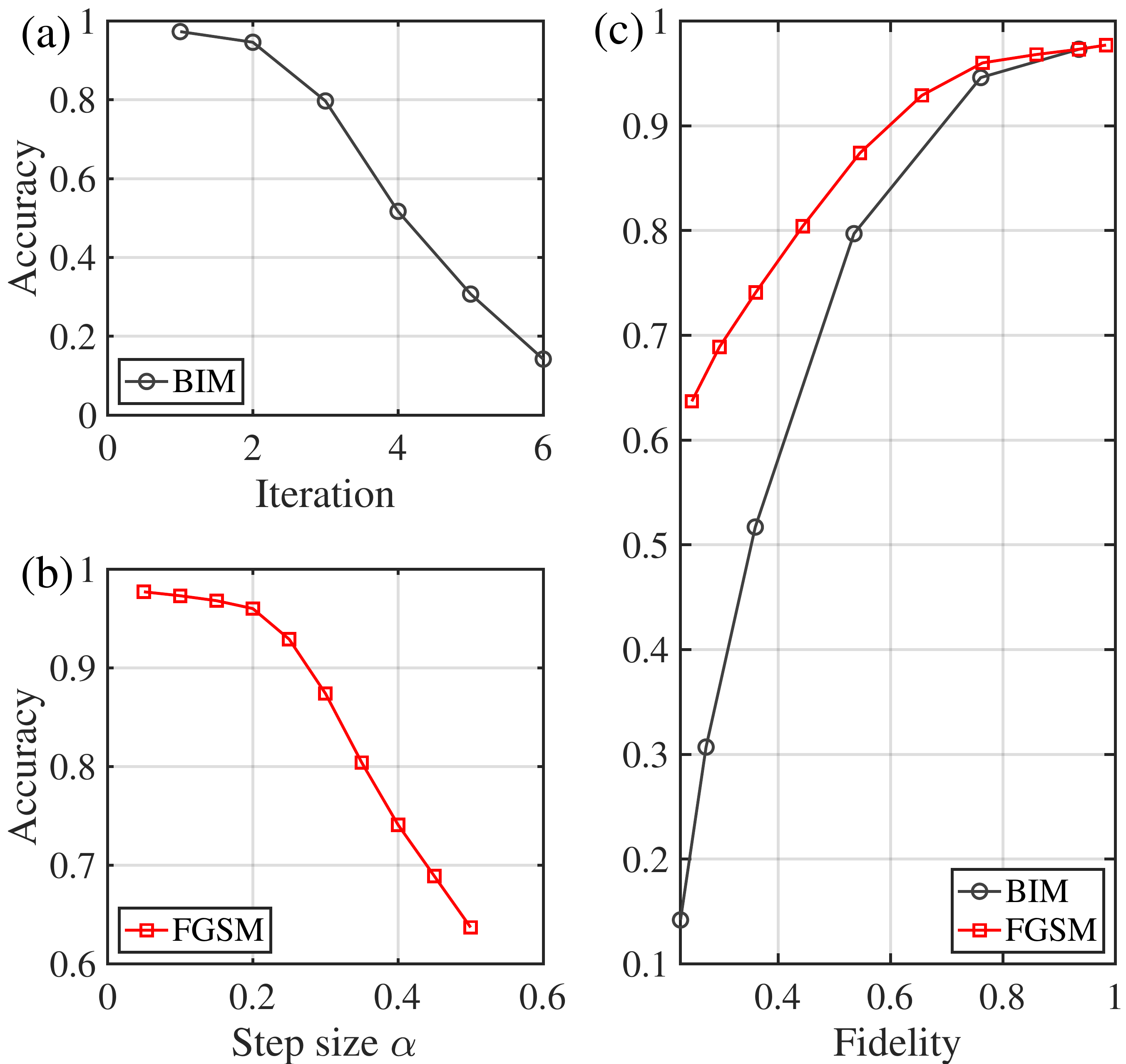}
\caption{Effects of adversarial untargeted functional attack on the accuracy of the quantum classifier for the problem of classifying handwritten digits 1 and 9. Here, the adversarial perturbation operators are assumed to be a layer of local unitary transformation. We use both the BIM method and the FGSM method to obtain adversarial examples. (a) For the BIM method, we generated adversarial perturbations using different number of iterations with the fixed step size 0.1. (b) For the FGSM method, we generate adversarial perturbations using different step sizes, and the accuracy drops accordingly with increasing step size.\label{fig:M2_LU_acc_fid}}
\end{figure}

Now, we have demonstrated how to obtained adversarial examples for the quantum classifiers by additive attacks, where each component of the data vectors are modified independently. In real experiments, to realize such adversarial examples with quantum devices might be challenging because this requires implementations of complicated global unitaries with very high precision. To this end, a more practical approach is to consider functional attacks, where the adversarial perturbation operators are implemented with a layer of local unitary transformations. In this case, the searching space is much smaller than that for the additive attacks, hence we may not be able to find the most efficient adversarial perturbations. Yet, once we find the adversarial perturbations, it could be much easier to realize such perturbations in the quantum laboratory. To study functional attacks, in our numerical simulations we consider adding a layer of local unitary transformations before sending the quantum states to the classifiers. We restrict that these local unitaries are close to the identity operators so as to keep the perturbations reasonably small. We apply both the BIM and FGSM methods to solve Eq. \eqref{Eq:WBunT} in the white-box untargeted setting. Partial of our results for the case of functional attacks are plotted in Fig. \ref{fig:M2_LU_acc_fid}. From this figure, it is easy to see that the performances of both the BIM and FGSM methods are a bit poorer than that for the case of additive attacks. For instance, in the case of functional attacks after six BIM iterations there is still a residue accuracy about $14\%$ [see Fig. \ref{fig:M2_LU_acc_fid}(a)], despite the fact that the average fidelity  has already decreased to $0.2$ [see Fig. \ref{fig:M2_LU_acc_fid}(c)]. This is in sharp contrast to the case of additive attacks, where five BIM iterations are enough to reduce the accuracy down to zero [see Fig. \ref{fig:M2_M_4_acc_fid}(a)] and meanwhile maintain the average fidelity larger than $0.73$ [see Fig. \ref{fig:M2_M_4_acc_fid}(b)].  The reduction of the performances for both methods is consistent with the fact that the searching space for functional attacks are much smaller than that for additive attacks.

\subsubsection{White-box attack: targeted}\label{white-box-attack-targeted}

Unlike in the case of untargeted attacks, in targeted attacks the attacker attempts to mislead the classifier to classify a data sample incorrectly into a specific targeted category.  A good example that manifestly showcases the importance of targeted attacks occurs in face recognition as well: in some situations the attacker may attempt to disguise her face inconspicuously to be recognized as an authorized user of a laptop or phone that authenticates users through face recognition. This type of attack has a particular name of impersonation attack in classical adversarial learning. It has been shown surprisingly in Ref. \cite{sharif2016accessorize} that physically realizable and inconspicuous impersonation attacks can be carried out by wearing a pair of carefully-crafted glasses designed for deceiving the state-of-the-art face recognition systems. In this subsection, we show that quantum classifiers are likewise vulnerable to targeted attacks in the white-box setting.

\begin{figure}
\centering
\includegraphics[width=0.48\textwidth]{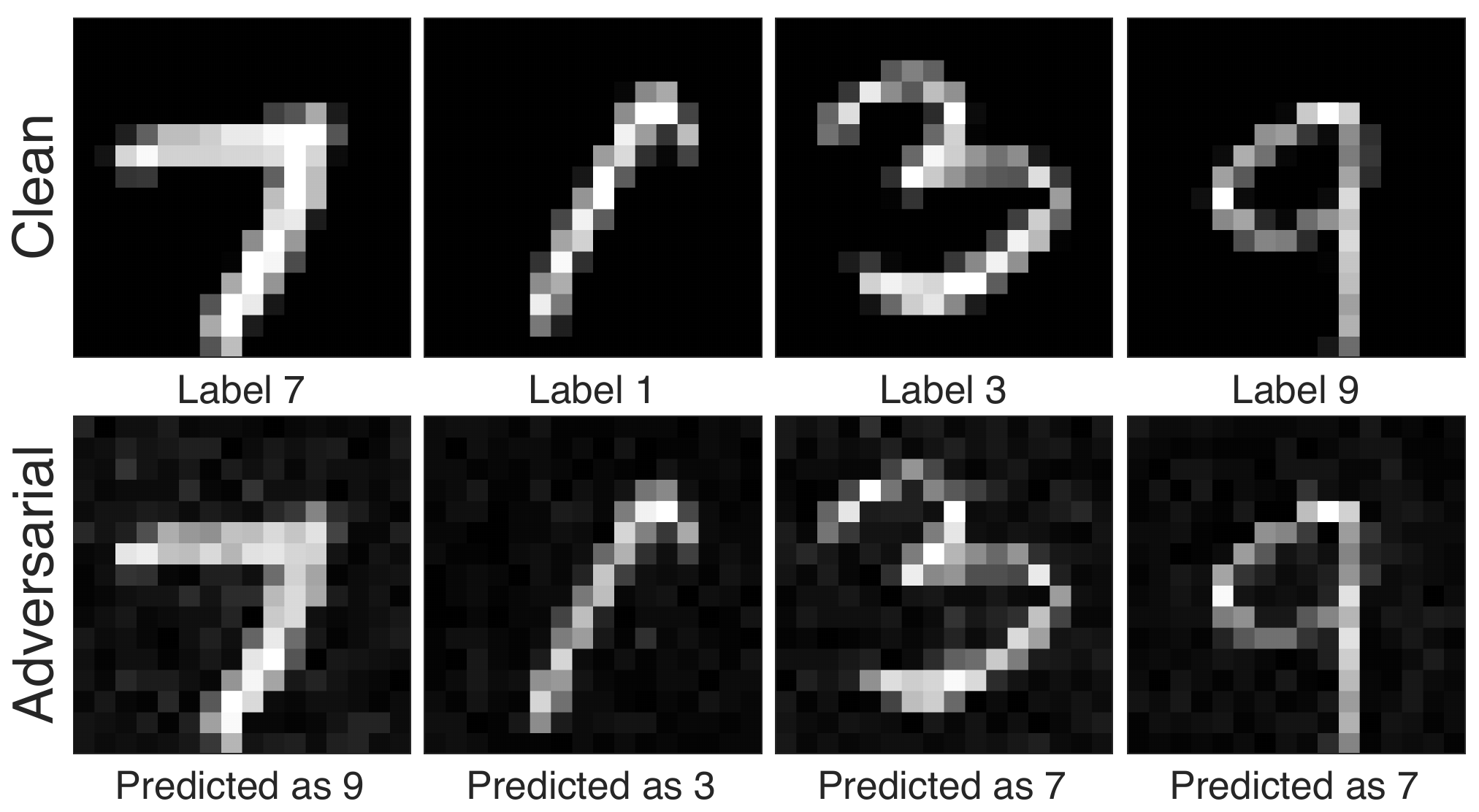}
\caption{\label{fig:adversarial-examples-targeted} Visual illustration of adversarial examples crafted using different attacks. From top to bottom: the clean and adversarial images generated for the quantum classifier by the BIM algorithm. By applying the additive attack, we can change the quantum classifier's classification result. The top images represent an correctly predicted legitimate example. The bottom images are incorrectly predicted adversarial example, even though they bear a close resemblance to the clean image. Here, the attacking algorithm we employed is BIM(0.1,3)}
\end{figure}

\begin{figure}
\centering
\includegraphics[width=0.48\textwidth]{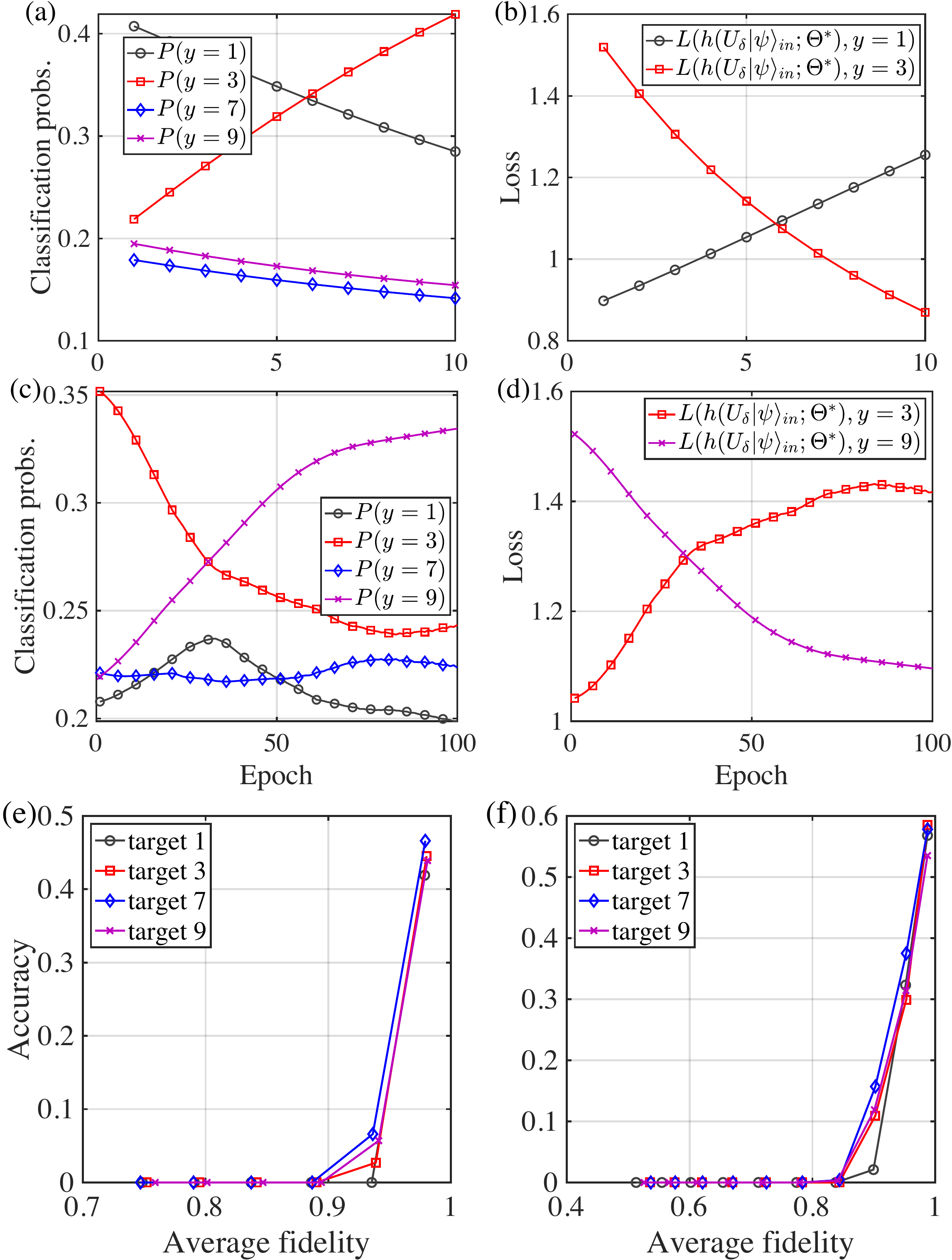}
\caption{\label{fig:targeted-attack} White-box targeted attacks for the four-category quantum classifier with depth $p=40$. (a) The classification probabilities for each digits as a function of the number of attacking epochs. Here, we use the BIM method to attack the quantum classifier. (b) The loss for classifying the image to be $1$ or $3$ as a function of the number of epochs. (c-d) Similar plots for the functional attacks.  (e-f) The accuracy as a function of the average fidelity during the attacking process. Here, we consider additive attacks with both the BIM (e) and FGSM (f) methods. 
}
\end{figure}

We consider attacking the four-category quantum classifier. In Fig. \ref{fig:adversarial-examples-targeted}, we randomly choose samples for digits $1$, $3$, $7$, and $9$ from MNIST and then solve the Eq. \eqref{eq:adversarial-learning-targeted} iteratively by the BIM method to obtain their corresponding adversarial examples. This figure shows the original legitimate images and their corresponding targeted adversarial ones for the four-category quantum classifier. For these legitimate samples, the quantum classifier can assign their labels correctly with high confidence.  But after targeted attacks, the same classifier is misled to classify the crafted images of digits $\{7,1,3, 9\}$ erroneously as the targeted digits $\{9,3,7,7\}$ with a decent high confidence, despite the fact that the differences between the crafted and legitimate images are almost imperceptible. To further illustrate how this works, in Figs. \ref{fig:targeted-attack} (a-d) we plot the classification probabilities for each digit and the loss functions with respect to particular digits as a function of the number of epochs. Here, we randomly choose an image of a given digit and then consider either additive [Figs.~\ref{fig:targeted-attack}(a-b)] or functional [Figs.~\ref{fig:targeted-attack}(c-d)] targeted attacks through the BIM method. For instance, in Fig.~\ref{fig:targeted-attack}(a) the image we choose is an image for digit $1$ and the targeted label is digit $3$. From this figure, at the beginning the quantum classifier is able to correctly identify this image as digit $1$ with probability $P(y=1)\approx 0.41$. As the number of BIM iteration increases $P(y=1)$ decreases and $P(y=3)$ increases, and after about six iterations $P(y=3)$ becomes larger than $P(y=1)$, indicating that the classifier begins to be deceived into predict the image as a digit $3$. Fig. \ref{fig:targeted-attack}(b) shows the loss as a function of the number of epochs. From this figure, as the iteration number increases, the loss for classifying the image as digit $1$ ($3$) increases (decreases), which is consistent with the classification probability behaviors in Fig. \ref{fig:targeted-attack}(a).

More surprisingly, 
we can in fact fool the quantum classifier to identify any images as a given targeted digit. This is clearly observed from Figs. \ref{fig:targeted-attack} (e-f) and Table \ref{table:performance-of-targeted-attack}, where we perform additive attacks for all the images of digits $\{1,3,7,9\}$ with different targeted labels and different attacking methods. In Figs. \ref{fig:targeted-attack}(e-f), we plot the accuracy versus the average fidelity. Here, for a given targeted label $l$ ($l=1,3,7,\text{or}, 9$), we perform additive attacks for all images with original labels not equal to $l$ and compute the accuracy and the average fidelity based on these images. From these figures, even when the average fidelity maintains larger than $0.85$ the accuracy can indeed decrease to zero, indicating that all the images are classified by the quantum classifier incorrectly as digit $l$. In Table \ref{table:performance-of-targeted-attack}, we summarize the performance of the BIM and FGSM methods in attacking the four-category quantum classifier in the white-box targeted setting.

\begin{table}
	\caption{The accuracy $\alpha^{\text{adv}}$ (in $\%$) and average fidelity $\bar{F}$ for the four-category quantum classifier with depth $p=10$ on the test dataset when being attacked by different methods for different targeted labels. Here, we consider additive attacks with both the BIM and FGSM methods. For the BIM method, we generate adversarial examples using three iterations with a step size of $0.05$. Whereas, for the FGSM method, we use a single step with step size of $0.03$. \label{table:performance-of-targeted-attack}}
	\begin{ruledtabular}
	\begin{tabular}{l|llllll}
		\diagbox{Attacks}{Targets} && 1 & 3 & 7 & 9  \\ 
		\colrule
		\multirow{2}{8em}{BIM(3, 0.05)} &$\alpha^{\text{adv}}$& \(5.7\%\) & \(6.6\%\) & \(2.7\%\) & \(0.0\%\) \\
		&\(\bar{F}\)&0.941 & 0.936 & 0.938 & 0.935\\
		\colrule
		\multirow{2}{8em}{FGSM(1, 0.03)}& $\alpha^{\text{adv}}$& \(2.1\%\) & \(10.9\%\)  & \(15.7\%\) & \(11.9\%\) \\
		&\(\bar{F}\)&0.899 & 0.902 & 0.902 & 0.901\\
	\end{tabular}
	\end{ruledtabular}
\end{table}

\subsubsection{Black-box attack:  transferability}\label{black-box-attack---transferability}

Unlike white-box attacks, black-box attacks assume limited or even no information about the internal structures of the classifiers and the learning algorithms. In classical adversarial learning, two basic premises that make black-box attacks possible have been actively studied \cite{li2018security}:  the \textit{transferability} of the adversarial examples and \textit{probing} the behavior of the classifier. Adversarial sample transferability is the property that an adversarial example produced to deceive one specific learning model can deceive another different model, even if their architectures differ greatly or they are trained on different sets of training data \cite{szegedy2013intriguing,goodfellow2014explaining,papernot2017practical}. Whereas, probing is another important premise of the black-box attack that the attacker uses the victim model as an oracle to label a synthetic training set for training a substitute model, hence  the attacker needs not even collect a training set to mount the attack. Here, we study the transferability of adversarial examples in a more exotic setting, where we first generate adversarial examples for different classical classifiers and then investigate whether they transfer to the quantum classifiers or not. This would have important future applications considering a situation where the attacker may only have access to classical resources.

Our results are summarized in Table \ref{table:transferability-table}. To obtain these results, we first train two classical classifiers, one based on a convolutional neural network (CNN) and the other based on a feedforward neural network (see Appendix. \ref{appendix} for details), with training data from the original MNIST dataset. Then we use three different methods (i.e., BIM, FGSM, and MIM) to produce adversarial examples in a white-box untargeted setting for both classical classifiers separately. After these adversarial examples are obtained, we evaluate the performance of the trained quantum classifier  on them.  From Table \ref{table:transferability-table}, it is evident that the performance of the quantum classifier on the adversarial examples is much worse than that on the original legitimate samples. For instance, for the adversarial examples generated for the CNN classifier by the MIM method, the accuracy of the quantum classifier is only $62.3\%$, which is $29.7\%$ lower than that for the clean legitimate samples. This  indicates roughly that $29.7\%$ of the adversarial examples originally produced for attacking the CNN classifier transfer to the quantum classifier. This transferability ratio may not be as large as that for adversarial transferability between two classical classifiers. Yet, given the fact that the structure of the quantum classifier is completely different from the classical ones, it is in fact a bit surprising that such a high transferability ratio can be achieved in reality. We expect that if we use another quantum classifier to play as the surrogate classifier, the transferability ratio might increase significantly. We leave this interesting problem for future studies.

\begin{table}
	\caption{\label{table:transferability-table}Black-box attacks to the quantum classifier. Here, the adversarial examples are generated by three different methods (i.e., BIM, FGSM, and MIM) for two different classical classifiers, one based on CNN and the other on FNN (see Appendix ). This table shows the corresponding accuracy (in $\%$) for each case on the MNIST test dataset. We denote the predication accuracy of the classical neural networks (quantum classifier) on the test set as $\alpha_C$ ($\alpha_Q$), and the predication accuracy on the adversarial test set as $\alpha_C^{\text{adv}}$ ($\alpha_Q^{\text{adv}}$).  The accuracy of the quantum classifier drops significantly on the adversarial examples generated for the classical neural networks.}
	\begin{ruledtabular}
	\begin{tabular}{cccccc}
		&\diagbox{{Attacks}}{{Accuracy}} & {$\alpha_C^{\text{adv}}$} & $\alpha_C-\alpha_C^{\text{adv}}$ & $\alpha_Q^{\text{adv}}$ & $\alpha_Q-\alpha_Q^{\text{adv}}$  \\ 
		\colrule
		\multirow{3}{2em}{CNN}&BIM (50, 0.01) & \(0.07\%\) & \(98.2\%\) & \(66.4\%\) & \(25.6\%\) \\
		&FGSM (1, 0.3) & \(0.6\%\) & \(98.3\%\) & \(51.6\%\) & \(40.4\%\) \\
		&MIM (10, 0.06) & \(0.7\%\) & \(98.2\%\) & \(62.3\%\) & \(29.7\%\)\\
		\colrule
		\multirow{3}{2em}{FNN}&BIM (50, 0.01) & \(0.6\%\) & \(99.3\%\) & \(68.1\%\) & \(23.9\%\)\\
		&FGSM (1, 0.3) & \(1.0\%\) & \(98.9\%\) & \(56.8\%\) & \(35.2\%\)\\ 
		&MIM (10, 0.06) & \(0.8\%\) & \(99.1\%\) & \(59.9\%\) & \(32.1\%\)\\
	\end{tabular} 
	\end{ruledtabular}
\end{table}

\subsubsection{Adversarial perturbations are not random noises}\label{adversarial-perturbations-are-not-noises}

The above discussions explicitly demonstrated the vulnerability of quantum classifiers against adversarial perturbations. The existence of adversarial examples is likewise a general property for quantum learning systems with high-dimensional Hilbert space. For almost all the images of hand-writing digits in MNIST, there always exists at least one corresponding adversarial example. Yet, it is worthwhile to clarify that adversarial perturbations are {\it not} random noises. They are carefully-engineered to mislead the quantum classifiers and in fact only occupy a tiny subspace of the total Hilbert space. To demonstrate this more explicitly, we compare the effects of random noises on the accuracy of both two- and four-category quantum classifiers with the effects of adversarial perturbations.   For simplicity and concreteness, we consider the uncorrelated decoherence noises  that occur in a number of experimental platforms (such as, Rydberg atoms, superconducting qubits, and trapped ions, etc.) for quantum computing \cite{Saffman_2016,krantz2019quantum,bruzewicz2019trapped,wu2018noise}:     
\begin{equation}
\mathcal{E}_{\text{depl}}(\rho)=(1-\beta)\rho+\frac{\beta}{3}(\sigma^x\rho\sigma^x +\sigma^y\rho\sigma^y +\sigma^z\rho\sigma^z ),\label{Eq:DepolNoise}
\end{equation}
where $\rho$ denotes the density state of a qubit, $\sigma^{x,y,z}$ are the usual Pauli matrices, and $\beta\in [0,1]$ is a positive number characterizing the strength of the decoherence noises.

\begin{figure}
\centering
\includegraphics[width=0.48\textwidth]{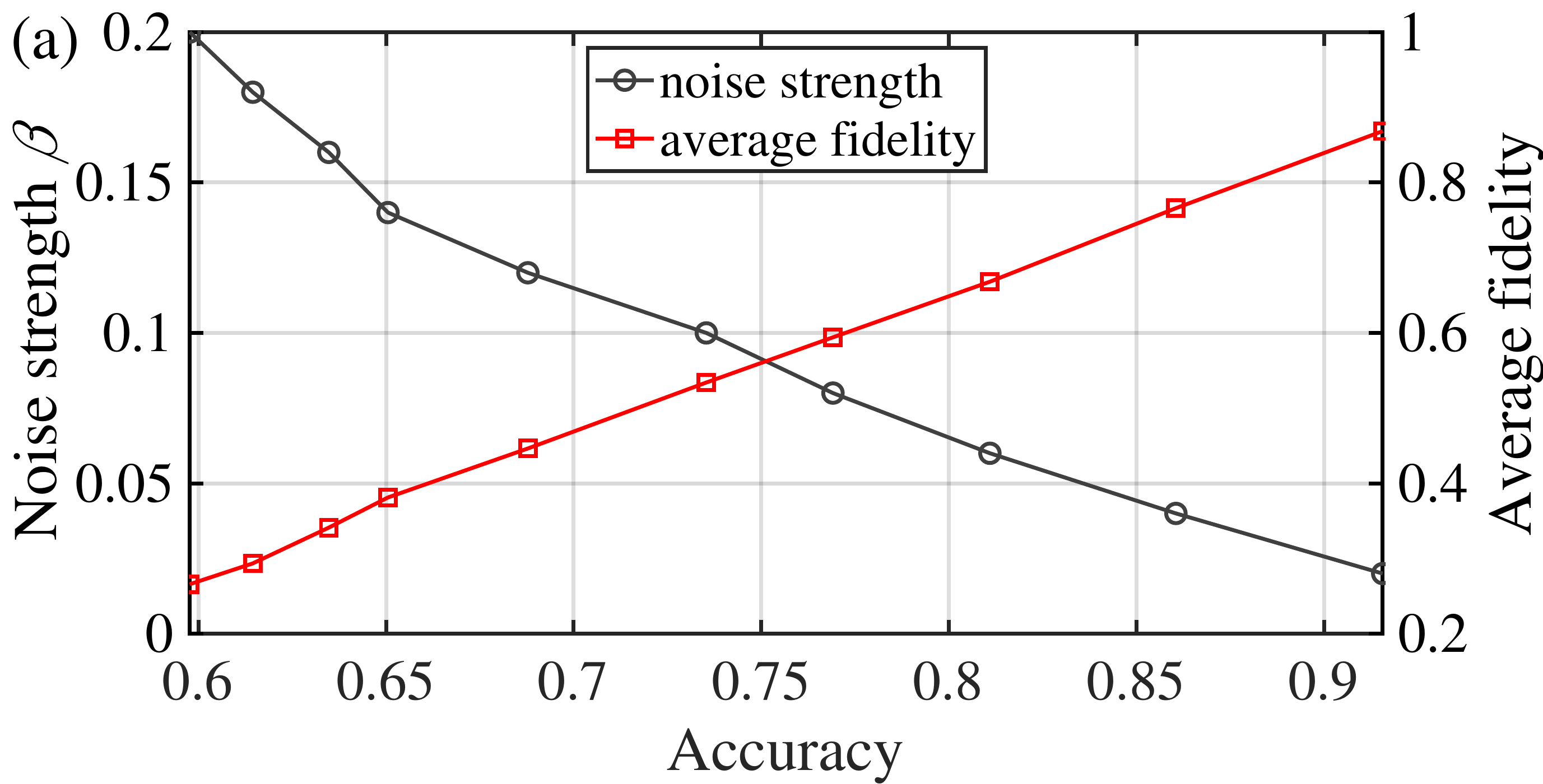}
\includegraphics[width=0.48\textwidth]{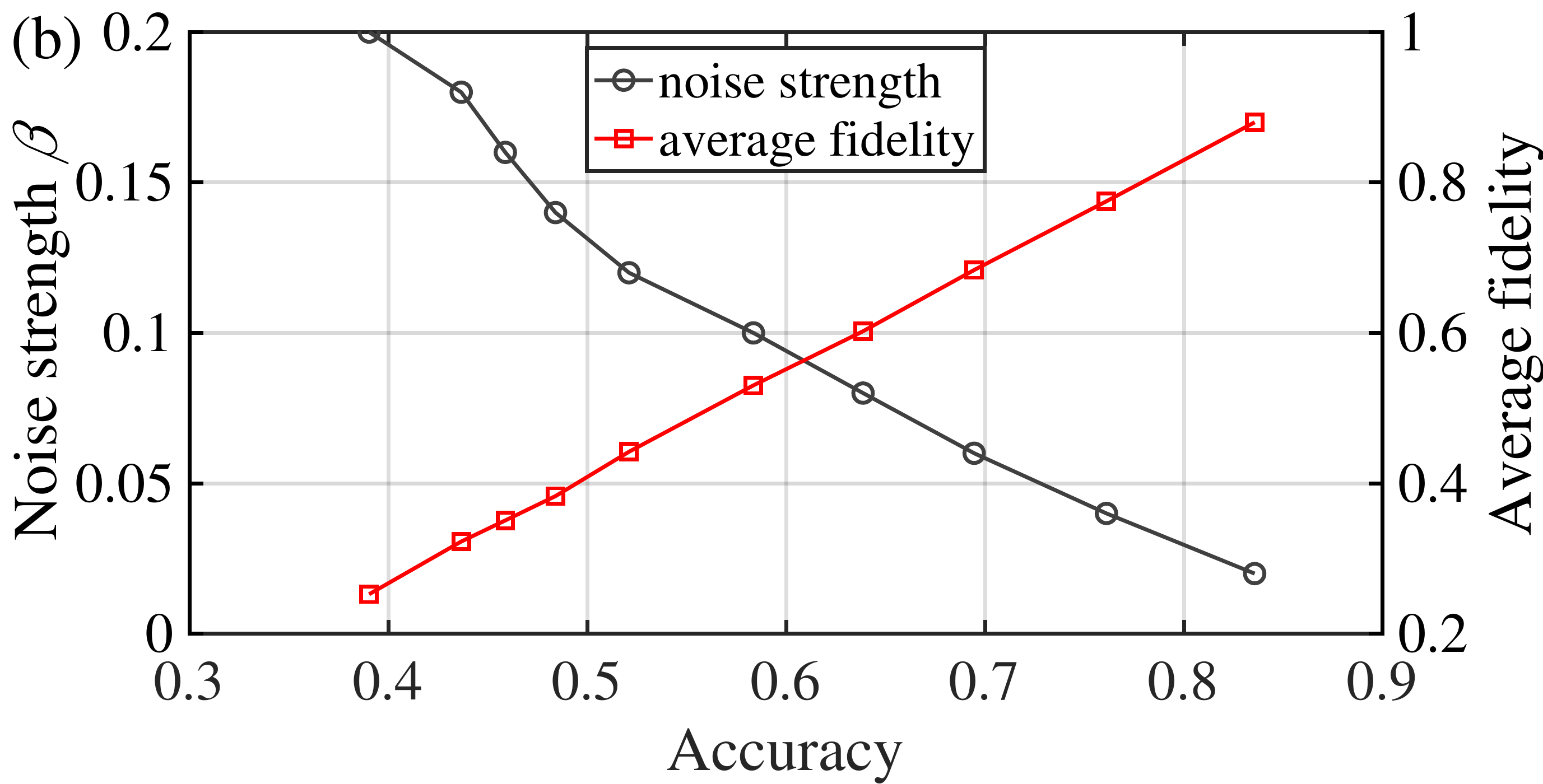}
\caption{\label{fig:noise_acc_p_fid} Effects of depolarizing noises with varying strength on the accuracy of the quantum classifiers with depth $p=20$. The mean classification accuracy is computed on the test set with respect to the fidelity between the original input states and the states affected by depolarizing noises on each qubit with varying strengths. The accuracy and fidelity are averaged over $1000$ random realizations. 
(a) Results for   the two-category quantum classifier. (b) Results for the four-category quantum classifier.}
\end{figure}

In Fig.~\ref{fig:noise_acc_p_fid}, we plot the  classification accuracy of the quantum classifiers versus the noise strength $p$ and the average fidelity between the original state and the state affected by a single layer of depolarizing noise on each qubit described by Eq. \ref{Eq:DepolNoise}. From this figure, we observe that the accuracy for both the two- and four-category quantum classifiers decreases roughly linearly with the increase of $p$ and the decrease of the average fidelity. This is in sharp contrast to the case for adversarial perturbations [see Fig. \ref{fig:targeted-attack} (e-f), Fig.~\ref{fig:M2_LU_acc_fid}(c), and Fig.~\ref{fig:M2_M_4_acc_fid}(b)(d) for comparison], where the accuracy has a dramatic reduction as the average fidelity begins to decrease from unity, indicating that the adversarial perturbations are not random noises. In fact, since the accuracy only decreases linearly with the average fidelity, this result also implies that quantum classifiers are actually rather robust to random noises. We mention that one may also consider the bit-flip or phase-flip noises and observe similar results. The fact that the adversarial perturbations are distinct from random noises is also reflected in our numerical simulations of the defense strategy by data augmentation---we find that the performance of the quantum classifier is noticeably better if we augment the training set by adversarial examples, rather than samples with random noises.

\subsection{Quantum adversarial learning topological phases of matter}\label{adversarial-learning-topological-phases-of-matter}

Classifying different phases and the transitions between them is one of the central problems in condensed matter physics. Recently, various machine learning tools and techniques have been adopted to tackle this intricate problem. In particular, a number of supervised and unsupervised learning methods have been introduced to classify phases of matter and identify phase transitions \cite{Wang2016Discovering,Zhang2016Triangular,Carrasquilla2017Machine,van2017Learning,Broecker2017Machine,Chng2017Machine,Wetzel2017Unsupervised,Hu2017Discovering,Hsu2018Machine,rodriguez2019identifying,Zhang2018Machine,Sun2018Deep,Huembeli2018Identifying}, giving rise to an emergent research frontier for machine learning phases of matter. Following these theoretical approaches, proof-of-principle experiments with different platforms \cite{Lian2019Machine,rem2019identifying,bohrdt2019classifying,zhang2019machine}, such as doped $\text{CuO}_2$ \cite{zhang2019machine}, electron spins in diamond nitrogen-vacancy centers \cite{Lian2019Machine}, and cold atoms in
optical lattices \cite{rem2019identifying,bohrdt2019classifying}, have been carried out in  laboratories to demonstrate their feasibility and unparalleled potentials. In addition, the vulnerability of these machine learning approaches to adversarial perturbations has  been pointed out in a recent work as well \cite{Jiang2019Vulnerability}.  It has been shown that typical phase classifiers based on classical deep neural networks are extremely vulnerable to adversarial attacks: adding a tiny amount of carefully-crafted noises or even just changing a single pixel of the legitimate sample may cause
the classifier to make erroneous predictions with a surprisingly high confidence level.

\begin{figure}
\centering
\includegraphics[width=0.48\textwidth]{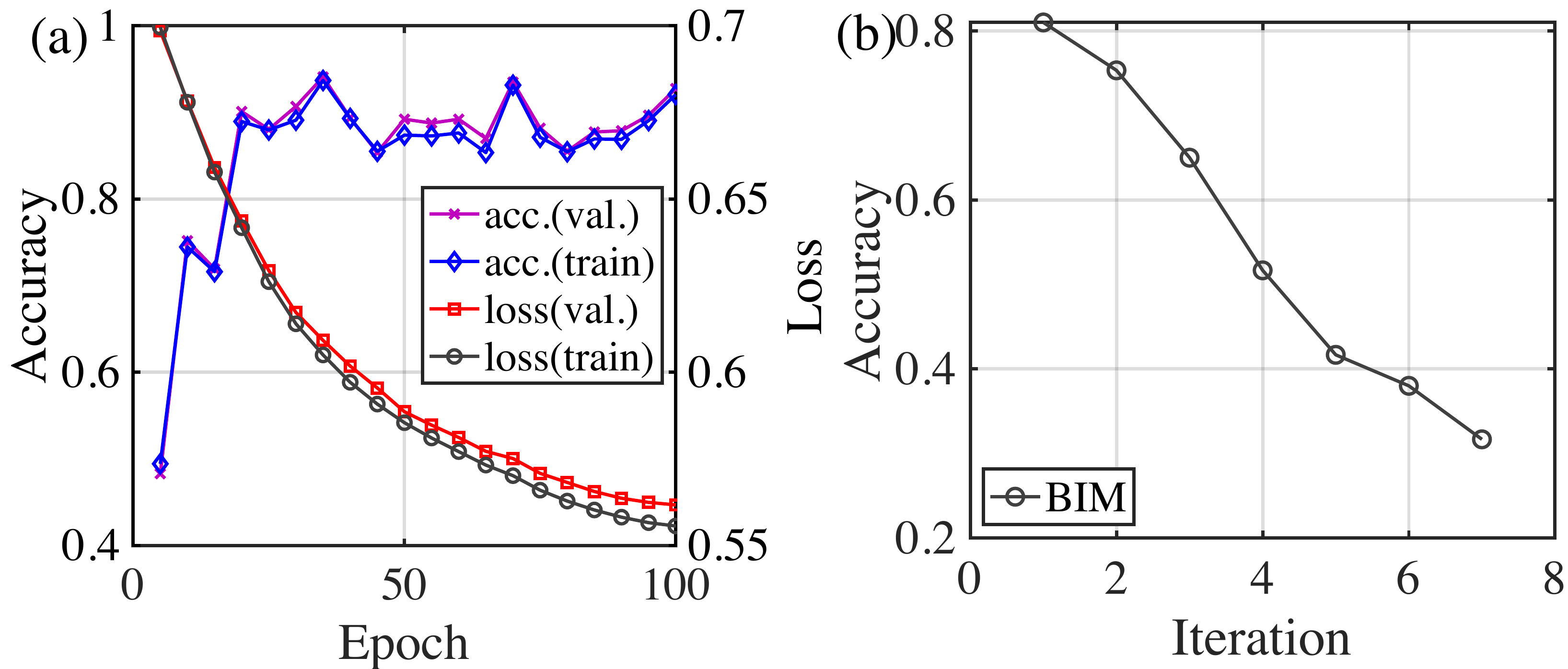}
\caption{\label{fig:TOF-train-attack} (a) The average accuracy and loss for the two-category quantum classifier as a function of the number of epochs. Here, we use a quantum classifier with structures shown in Fig.~\ref{CircuitQC} and depth ten ($p=10$) to  perform binary classification for topological/non-topological phases. To train the classifier, we use the Adam optimizer with a batch size of $512$ and a learning rate of $0.005$ to minimize the loss function in Eq.~\eqref{eq:loss}.  The accuracy and loss are averaged on $19956$ training samples and $6652$ validation samples. (b) The accuracy of the quantum classifier as a function of the iterations of the BIM attack. Here, the BIM step size is 0.01.}
\end{figure}

\begin{figure}
\centering
\includegraphics[width=0.48\textwidth]{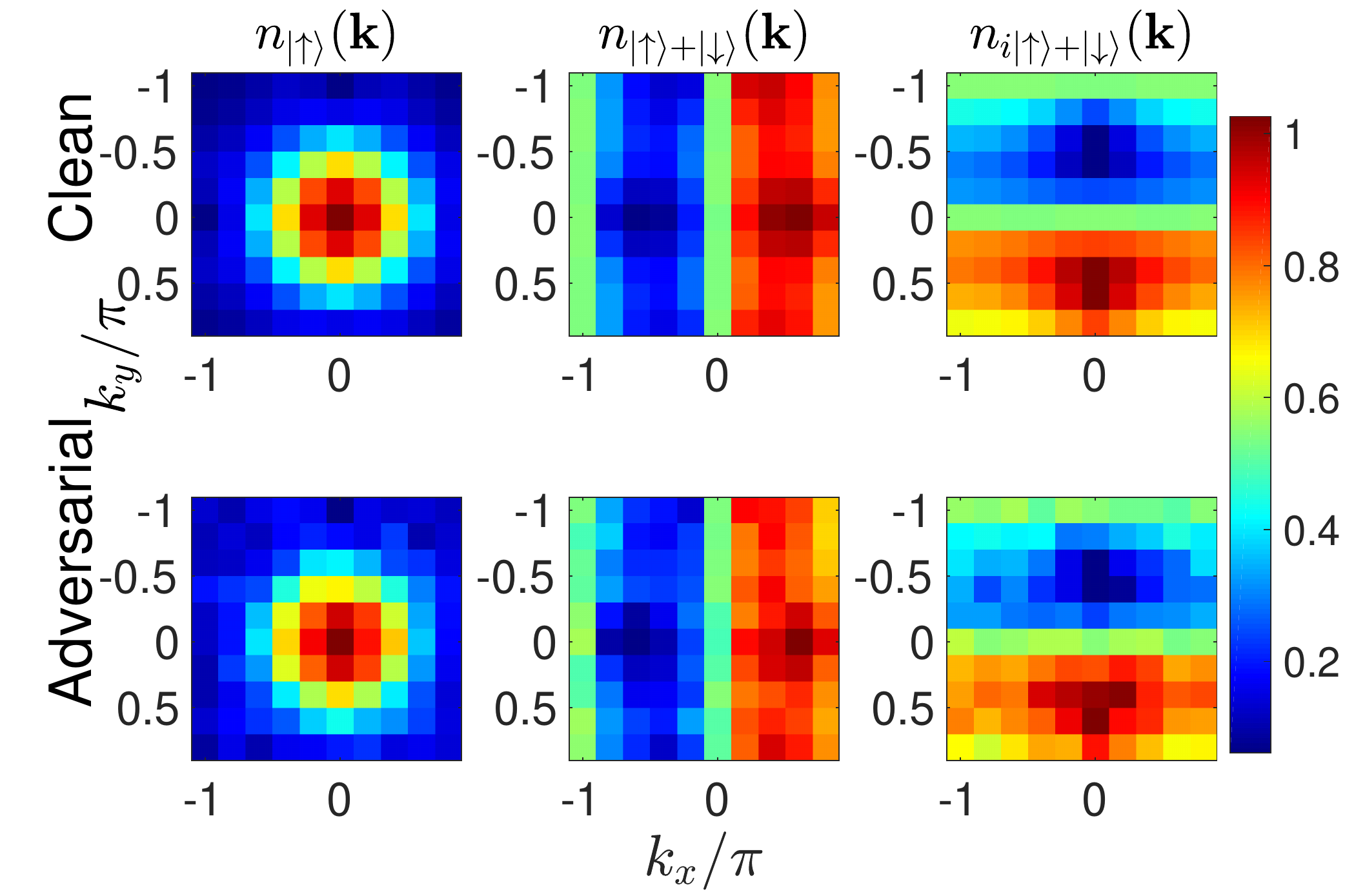}
\caption{\label{fig:TOF-2D-adversarial} The clean and the corresponding adversarial time-of-flight images for using the quantum classifier to classify topological phases. (Top) A legitimate sample of the density distribution in momentum space for the lower band with lattice size \(10\times 10\). (Bottom) An adversarial example obtained by the fast gradient sign method, which only differs with the original one by a tiny amount of noises that are imperceptible to human eyes.}
\end{figure}

Despite these exciting progresses made in the area of machine learning phases of matter, most previous approaches are based on classical classifiers and using quantum classifiers to classify different phases and transitions still remains barely explored hitherto. Here, in this section we study the problem of using quantum classifiers to classify different phases of matter, with a focus on topological phases that are widely believed to be more challenging than conventional symmetry-breaking phases (such as the paramagnetic/ferromagnetic phases) for machine-learning approaches \cite{Zhang2016Triangular,Zhang2017Machine,Zhang2018Machine,Sun2018Deep}. We show, through a concrete example, that the quantum classifiers are likewise vulnerable to adversarial perturbations. We consider the following 2D square-lattice model for quantum anomalous Hall (QAH) effect, where a combination of spontaneous magnetization and spin-orbit coupling leads to quantized Hall conductivity in the absence of an external magnetic field: 
\begin{eqnarray}
H_{\text{QAH}}  &=&  J_{\text{SO}}^{(x)}\sum_{\mathbf{r}}[(c_{\mathbf{r}\uparrow}^{\dagger}c_{\mathbf{r}+\hat{x}\downarrow}-c_{\mathbf{r}\uparrow}^{\dagger}c_{\mathbf{r}-\hat{x}\downarrow})+\text{H.c.}] \label{eq:Ham-QAH}
\\
&+& iJ_{\text{SO}}^{(y)}\sum_{\mathbf{r}}[(c_{\mathbf{r}\uparrow}^{\dagger}c_{\mathbf{r}+\hat{y}\downarrow}-c_{\mathbf{r}\uparrow}^{\dagger}c_{\mathbf{r}-\hat{y}\downarrow})+\text{H.c.}] \nonumber \\
 &-& t\sum_{\langle\mathbf{r},\mathbf{s}\rangle}(c_{\mathbf{r}\uparrow}^{\dagger}c_{\mathbf{s}\uparrow}-c_{\mathbf{r}\downarrow}^{\dagger}c_{\mathbf{s}\downarrow})+\mu\sum_{\mathbf{r}}(c_{\mathbf{r}\uparrow}^{\dagger}c_{\mathbf{r}\uparrow}-c_{\mathbf{r}\downarrow}^{\dagger}c_{\mathbf{r}\downarrow}). \nonumber
\end{eqnarray}
Here $c_{\mathbf{r}\sigma}^{\dagger}$ $(c_{\mathbf{r}\sigma})$ is the fermionic creation (annihilation) operator with pseudospin $\sigma=(\uparrow,\downarrow)$ at site $\mathbf{r}$, and $\hat{x},\hat{y}$ are unit lattice vectors along the $x,y$ directions. The first two terms describe the spin-orbit coupling with $J_{\text{SO}}^{(x)}$ and $J_{\text{SO}}^{(y)}$ denoting its strength along the $x$ and $y$ directions, respectively. The third and the fourth terms denote respectively the spin-conserved nearest-neighbor hopping and the on-site Zeeman interaction. In momentum space, this Hamiltonian has two Bloch bands and the topological structure of this model can be characterized by the first Chern number:
\begin{equation}
    C_1=-\frac{1}{2\pi}\int_{\text{BZ}}d k_x dk_y F_{xy}(\mathbf{k}), \label{ChernN}
\end{equation}
where $F_{xy}$ denotes the Berry curvature $F_{xy}(\mathbf{k})\equiv\partial_{k_{x}}A_{y}(\mathbf{k})-\partial_{k_{y}}A_{x}(\mathbf{k})$ with the Berry connection $A_{\mu}(\mathbf{k})\equiv\langle \varphi(\mathbf{k})|i\partial_{k_{\mu}}|\varphi(\mathbf{k})\rangle$ [$\mu=x,y$ and $\varphi(\mathbf{k})$ is the Bloch wavefunction of the lower band], and the integration is over the whole first Brillouin zone (BZ). It is straightforward to obtain that $C_1=-\sign(\mu)$ when $0<|\mu|<4t$ and $C_1=0$ otherwise.

\begin{table}
	\caption{Average fidelity $\bar{F}$ and accuracy (in \(\%\)) of the two-category quantum classifier with depth $p=10$ when being attacked by the BIM and FGSM methods in the white-box untargeted setting. Here, the accuracy and fidelity are averaged over 2000 testing samples. \label{table:TOF-performance-of-untargeted-attack}}
	\begin{ruledtabular}
	\begin{tabular}{l l l}
		Attacks & $\bar{F}$ & Accuracy  \\ 
		\colrule
		BIM (3, 0.01) & 0.988 & \(31.6\%\) \\ 
		FGSM (1, 0.03) & 0.952 & \(6.3\%\) \\ 
	\end{tabular} 
	\end{ruledtabular}
\end{table}

The above Hamiltonian can be implemented with synthetic spin-orbit couplings in cold-atom experiment \cite{Liu2014Realization} and the topological index $C_1$ can be obtained from the standard time-of-flight images \cite{alba2011seeing,deng2014direct}. Indeed, by using ultracold fermionic atoms in a periodically modulated optical honeycomb lattice, the experimental realization of the Haldane model, which bears similar physics and Hamiltonian structures as in Eq. \eqref{ChernN}, has been reported \cite{jotzu2014experimental}. For our purpose, we first train a two-category quantum classifier to assign labels of $C_1=0$ or $C_1=1$ to the time-of-flight images. To obtain the training data, we 
diagonalize the Hamiltonian in Eq. \eqref{eq:Ham-QAH} with an open boundary condition and calculate the atomic density distributions with different spin bases for the lower band. These density distributions can be directly measured through the time-of-flight imaging techniques in cold atom experiments and serve as our input data. We vary $\lambda_{\text{SO}}$ and $t$ in both the topological and topologically trivial regions to generate several thousand of data samples. Similar as in the above discussion on identifying images of hand-writing digits, we use amplitude encoding to convert the data for density distributions to the input quantum states for the quantum classifier.  In Fig.~\ref{fig:TOF-train-attack}(a), we plot the average accuracy and loss as a function of the number of epochs. It shows that after training, the quantum classifier can successfully identify the time-of-flight images with reasonably high accuracy. Yet, we note that this accuracy is a bit lower than that for the case of classifying paramagnetic/ferromagnetic phases discussed in the next section, which is consistent with the general belief that topological phases are harder to learning.

Unlike the conventional phases or the hand-writing digit images, topological phases are described by nonlocal topological invariants (such as the first Chern number), rather than local order parameters. Thus, intuitively the obtaining of adversarial examples might also be more challenging, since the topological invariants capture only the global properties of the systems and are insensitive to local perturbations. Yet, here we show that adversarial examples do exist in this case and the quantum classifier is indeed vulnerable in learning topological phases. To obtain adversarial examples, we consider attacking the quantum classifier additively in the white-box untargeted setting. Partial of our results are plotted in Fig.~\ref{fig:TOF-train-attack}(b). From this figure, the accuracy for the quantum classifier in classifying time-of-flight images decreases rapidly as the number of attacking iterations increases and after about six iterations it becomes less than $0.4$, indicating that more than $60\%$ the attacked images in the test set are misclassified. To illustrate this even more concretely, in Fig.\ref{fig:TOF-2D-adversarial} we randomly choose a time-of-flight image and then solve the Eq. \eqref{Eq:WBunT} iteratively by the BIM method to obtain its corresponding adversarial examples. Again, as shown in this figure the obtained adversarial example looks like the same as the clean legitimate time-of-flight image. They differ only by a tiny amount of perturbation that is imperceptible to human eyes. In addition, we summarize the performance of two different methods (BIM and FGSM) in attacking the quantum classifier in Table \ref{table:TOF-performance-of-untargeted-attack}. Both the BIM and FGSM methods perform noticeably well.

\begin{figure}
\centering
\includegraphics[width=0.48\textwidth]{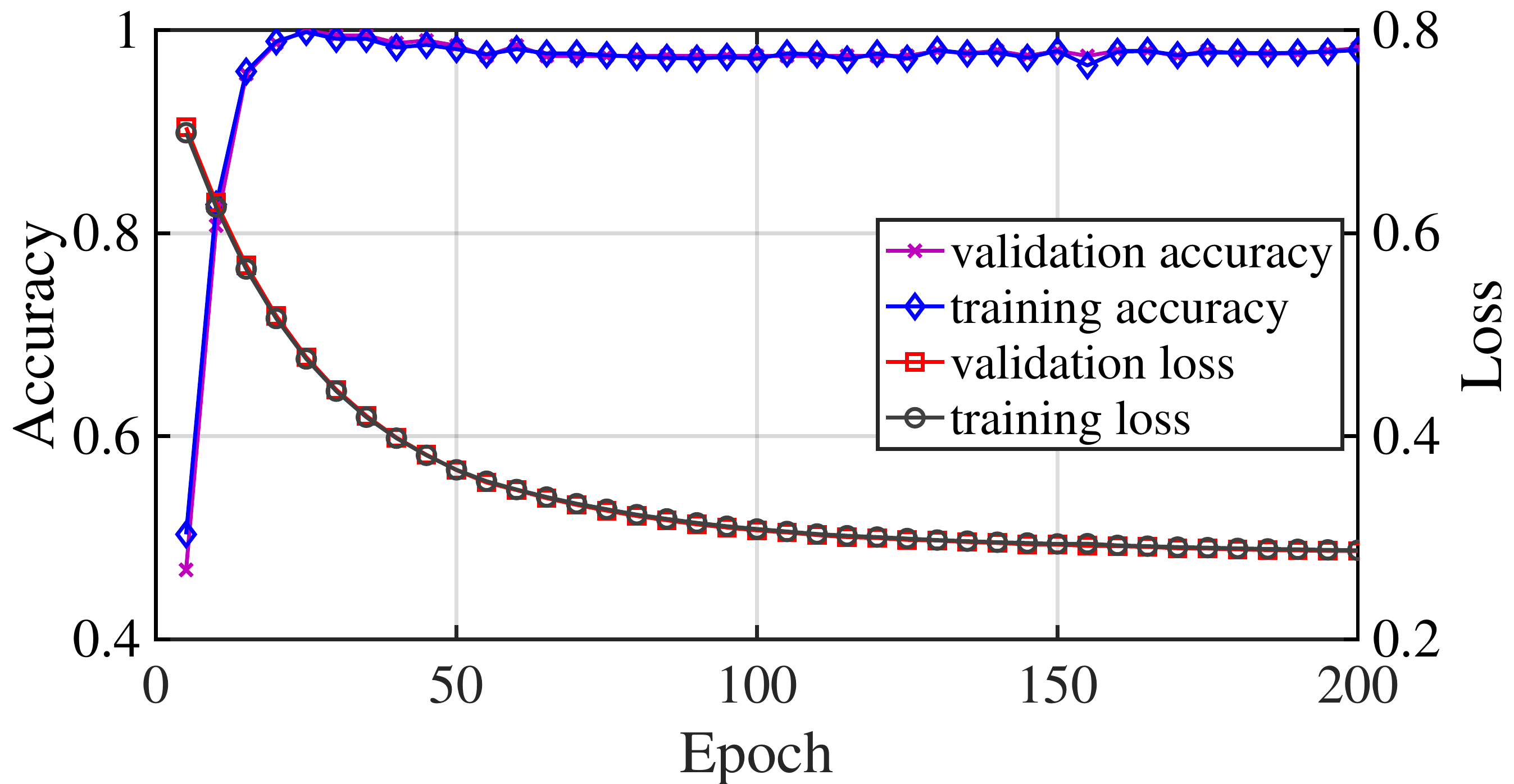}
\caption{\label{fig:tf1d-train} The average accuracy and loss function as a function of the number of training steps. We use a depth-10 quantum classifier with structures shown in  Fig.~\ref{CircuitQC} to classify the ferromagnetic/paramagnetic phases for the ground states of $H_{\text{Ising}}$. We plot the accuracy of $1182$ training samples and $395$ validation samples (which are not in the training dataset). We present the results of the first $200$ iteration epochs. The learning rate is 0.005. The difference between the training loss and validation loss is very small, indicating that the quantum classifier does not overfit. The final accuracy on the $395$ test samples is roughly (98\%).
}
\end{figure}

\subsection{Adversarial learning quantum data}\label{adversarial-learning-quantum-data}

In the above discussion, we considered using quantum classifiers to classify classical data (images) and studied their vulnerabilities to adversarial perturbations. This may have important applications in solving practical machine learning problems in our daily life. However, in such a scenario a prerequisite is to first transfer classical data to quantum states, which may require certain costly processes or techniques (such as quantum random access memories \cite{giovannetti2008quantum}) and thus renders the potential quantum speedups nullified \cite{Aaronson2015Read}. Unlike classical classifiers that can only take classical data as input, quantum classifiers can also classify directly quantum states produced by quantum devices. Indeed, it has been shown that certain quantum classifiers, such as quantum principal component analysis \cite{Lloyd2014Quantum} and quantum support vector machine \cite{rebentrost2014quantum}, could offer an exponential speedup over their classical counterparts in classifying quantum data directly. In this subsection, we consider the vulnerability of quantum classifiers in classifying quantum states.

For simplicity and concreteness, we consider the following 1D transverse field Ising model:
\begin{equation}
H_{\text{Ising}}=-\sum_{i=1}^{L-1}\sigma_i^z\sigma_{i+1}^z-J_x\sum_{i=1}^{L}\sigma_i^x,
\label{eq:Ham-Ising}
\end{equation}
where $\sigma_i^z$ and $\sigma_i^x$ are the usual Pauli matrices acting on the $i$-th spin and $J_x$ is a positive parameter describing the strength of the transverse field.   This model maps to free fermions through a Jordan–Wigner transformation and is exactly solvable. At zero temperature, it features a well-understood quantum phase transition at $J_x=1$, between a paramagnetic phase for $J_x>1$ and a ferromagnetic phase for $J_x<1$. It is an exemplary toy model for studying quantum phase transitions and an excellent testbed for different new methods and techniques. Here, we use a quantum classifier, with structures shown in Fig. \ref{fig:circuitclassifier},  to classify the ground states of $H_{\text{Ising}}$ with varying $J_x$ (from $J_x=0$ to $J_x=2$) and show that this approach is extremely vulnerable to adversarial perturbations as well.

\begin{figure}
\centering
\includegraphics[width=0.48\textwidth]{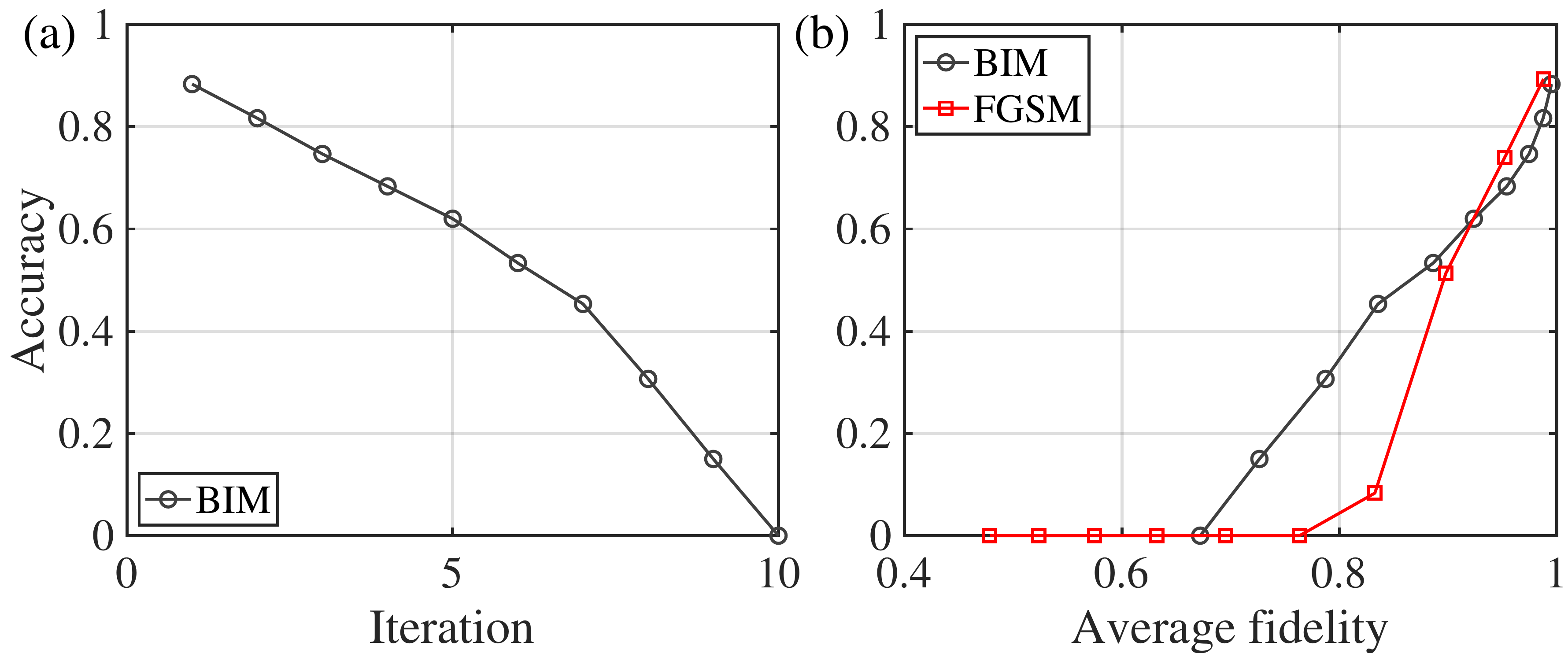}
\caption{\label{fig:TF1D_acc_fid} Effect of additive adversarial attack on the accuracy of the two-category quantum classifier in classifying the ferromagnetic/paramagnetic phases for the ground states of the transverse field Ising model. We use both the BIM and FGSM methods to generate adversarial examples in the white-box untargeted setting.  For the BIM method, we fix the step size to be $0.05$ and the iteration number to be ten. For the FGSM method, we perform the attack using a single step but with step size ranging from $0.1$ to $1.0$.  The circuit depth of the quantum classifier being attacked is $p=10$ and the system size for the Ising model is $L=8$. (a) The results for the BIM attack. (b) The accuracy as a function of average fidelity between the legitimate and adversarial samples for both the BIM and FGSM methods.}
\end{figure}

To generate the data sets for training, validation, and testing, we sample a series of Hamiltonians with varying $J_x$ from $0$ to $2$ and calculating their corresponding ground states, which are used as input data to the quantum classifier. We train the quantum classifier with the generated training dataset and our results for training is shown in Fig. \ref{fig:tf1d-train}. Strikingly, our quantum classifier is very efficient in classifying these ground states of $H_{\text{Ising}}$ into categories of paramagnetic/ferromagnetic phases and we find that a model circuit with depth $p=5$ is enough to achieve near-perfect classification accuracy. This is in contrast to the case of learning topological phases, where a quantum classifier with depth $p=10$ only gives an accuracy of around $90\%$. In addition, we mention that one can also use the quantum classifier to study the quantum phase transition.

Similar to the cases for classical input data, the quantum classifiers are vulnerable to adversarial perturbations in classifying quantum data as well. To show this more explicitly, we consider attacking the above quantum classifier trained with quantum inputs additively in the white-box untargeted setting. Partial of our results are plotted in Fig.~\ref{fig:TF1D_acc_fid}. In Fig. ~\ref{fig:TF1D_acc_fid}(a), we plot the accuracy as a function of the number of the BIM iterations and find that it decreases to zero after ten BIM iterations, indicating that all the slightly-adjusted quantum states, including even these far away from the phase transition point, are misclassified by the quantum classifier. In Fig. ~\ref{fig:TF1D_acc_fid}(b), we plot the accuracy as a function of averaged fidelity for different attacking methods.  From this figure, both the BIM and FGSM methods are notably effective in this scenario and the accuracy of the quantum classifier  on the generated adversarial examples decreases to zero, whereas the average fidelity maintains moderately large for both methods.

\section{Defense: quantum adversarial training}\label{defense-1}

In the above discussions, we have explicitly shown that quantum classifiers are vulnerable to adversarial perturbations. This may raise serious concerns about the reliability and security of quantum learning systems, especially for these applications that are safety and security-critical, such as self-driving cars and biometric authentications.  Thus, it is of both fundamental and practical importance to study possible defense strategies to increase the robustness of quantum classifiers to adversarial perturbations. 

In general, adversarial examples are hard to defend against because of the following two reasons. First, it is difficult to build a precise theoretical model for the adversarial example crafting process. This is a highly non-linear and non-convex sophisticated optimization process and we lack proper theoretical tools to analyse this process, making it notoriously hard to obtain any theoretical argument that a particular defense strategy will rule out a set of adversarial examples. Second, defending adversarial examples requires the learning system to produce proper outputs for every possible input, the number of which typically scales exponentially with the size of the problem. Most of the time, the machine learning models work very well but only for a very small ratio of all the possible inputs. Nevertheless, in the field of classical adversarial machine learning, a variety of defense strategies have been proposed in recent years to mitigate the effect of adversarial attacks, including adversarial training \cite{kurakin2016adversarial}, gradient hiding \cite{tramer2017space}, defensive distillation \cite{papernot2016distillation}, and defense-GAN \cite{samangouei2018defense}, etc.  Each of these strategies has its own advantages and disadvantages and none of them is adaptive to all types of adversarial attacks. In this section, we study the problem of how to increase the robustness of quantum classifiers against adversarial perturbations. We adopt one of the simplest and effective methods, namely adversarial training, to the case of quantum learning and show that it can significantly enhance the performance of quantum classifiers in defending adversarial attacks.

The basic idea of adversarial training is to strengthen model robustness by injecting adversarial examples into the training set. It is a straightforward brute force approach where one simply generates a lot of adversarial examples using one or more chosen attacking strategies and then retrain the classifier with both the legitimate and adversarial samples. For our purpose, we employ a \textit{robust optimization} \cite{ben2009robust} approach and reduce the task to solving a typical min-max optimization problem:
\begin{equation}\label{eq:adv-training-min-max}
    \min_{\Theta} \frac{1}{N} \sum_{i=1}^N \max_{U_{\delta}\in \Delta} L(h(U_{\delta}|\psi\rangle_{\text{in}}^{(i)});\Theta), y^{(i)}),
\end{equation}
where $|\psi\rangle_{\text{in}}^{(i)}$ is the $i$-th sample under attack, and $y^{(i)}$ denotes its original corresponding label. The meaning of Eq. \eqref{eq:adv-training-min-max} is clear: we are training the quantum classifier to minimize the adversarial risk, which is described by the average loss for the worst-case perturbations of the input samples. We mention that this min-max formulation has already been extensively studied in the field of robust optimization and many methods for solving such min-max problems have been developed \cite{ben2009robust}. One efficient method is to split Eq. \eqref{eq:adv-training-min-max} into two parts: the outer minimization and the inner maximization.  The inner maximization problem is exactly the same problem of generating adversarial perturbations, which have discussed in detail in Sec.~\ref{CALQC} and Sec.~\ref{VQC}. The outer minimization task boils down to a task of minimizing the loss function on adversarial examples. With this in mind, we develop a three-step procedure to solve the total optimization problem.  In the first step, we randomly choose a batch of input samples \({|\psi\rangle}_{\text{in}}^{(i)}\) together with their corresponding labels \(y^{(i)}\). Then, we calculate the `worst-case' perturbation of \({|\psi\rangle}_{\text{in}}^{(i)}\) with respect to the current model parameters \(\Theta_t\). That is to solve: \(U_{\delta^*}=\argmax_{U_\delta\in\Delta} L(h(U_{\delta}|\mathbf{\psi}\rangle;\Theta),y^{(i)})\). In the third step, we update the parameters \(\Theta_t\) according to the minimization problem at \(U_{\delta^*}{|\psi\rangle}_{\text{in}}\): \(\Theta_{t+1}=\Theta_{t}-\eta \nabla_{\Theta} L\left(h({U_{\delta^*}|\psi\rangle}_{\text{in}}^{(i)};\Theta_{t}), y^{(i)}\right)\). We repeat these three steps until the accuracy converges to a reasonable value.

\begin{figure}
\centering
\includegraphics[width=0.48\textwidth]{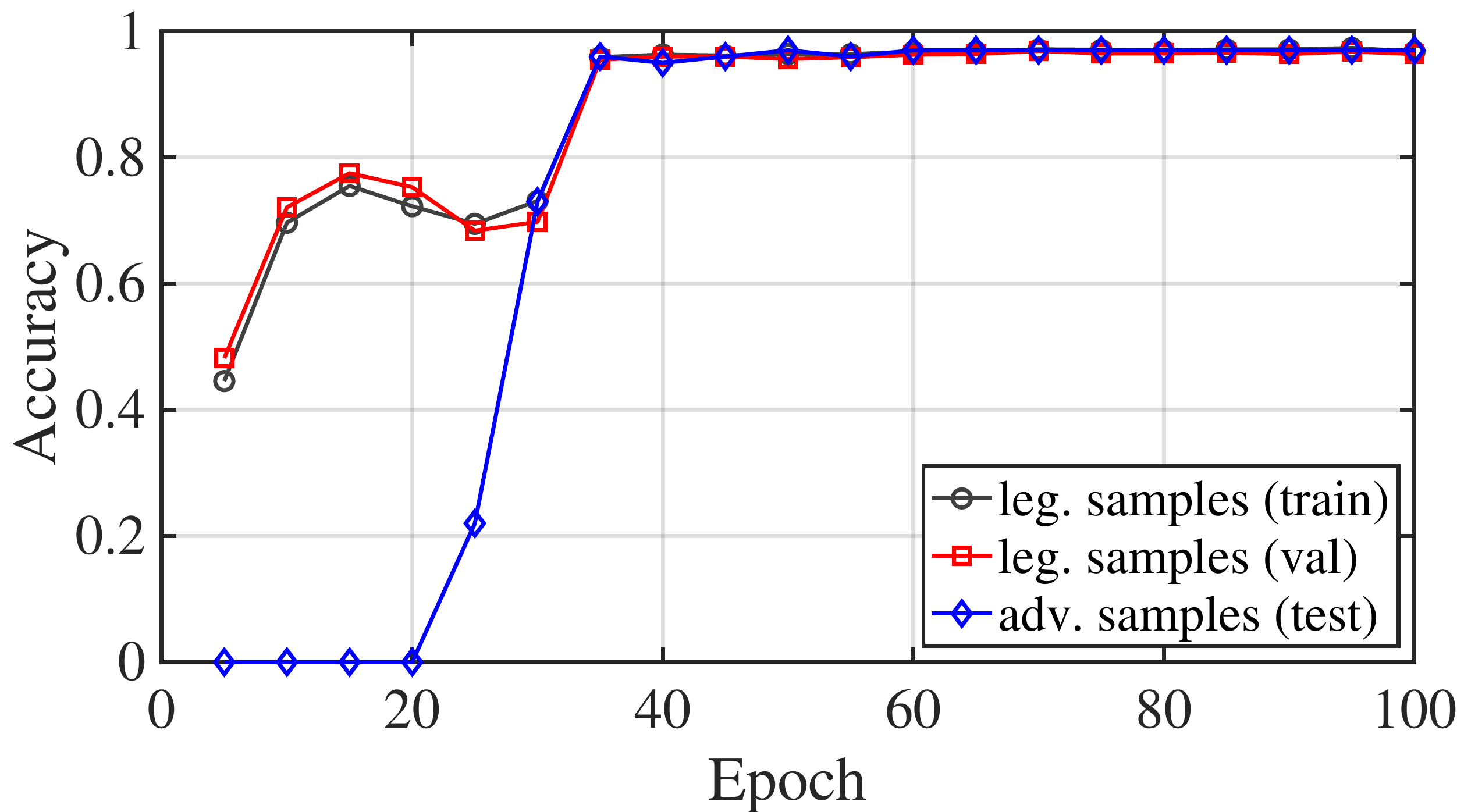}
\caption{\label{fig:Adversarial-Training} Strengthening the robustness of the quantum classifier against adversarial perturbations by quantum adversarial training. 
In each epoch, we first  generate adequate adversarial examples with the BIM method for the quantum classifier with the current model parameters. The iteration number is set to be three and the BIM step size is set to be $0.05$. Then, we train the quantum classifier with both the legitimate and crafted samples. The circuit depth of the quantum classifier is ten and the learning rate is set to be $0.005$.}
\end{figure}

Partial of our results are shown in Fig. \ref{fig:Adversarial-Training}. In this figure, we consider the adversarial training of a quantum classifier in identifying handwritten digits in MNIST. We use the BIM method in the white-box untargeted setting to generate adversarial examples. We use $20000$ clean images and generate their corresponding adversarial images. The clean images and the adversarial ones together form the training data set, and another 2000 images are used for the testing. From this figure, it is evident that, after adversarial training, the accuracy of the quantum classifier for both the adversarial samples and legitimate samples increases significantly. At the beginning of the training, the accuracy for the adversarial samples in the testing set remains zero. This is because the initial model parameters are randomly chosen, so the quantum classifier does not learn enough information and its performance on even legitimate samples is still very poor at the beginning (hence for each sample it is always possible to find an adversarial example by the BIM method, resulting in a zero accuracy on the testing set of adversarial examples). After the early stage of the adversarial training, this accuracy begins to increase rapidly and the quantum classifier is able to classify more and more crafted samples correctly. In other words, the BIM attack becomes less and less effective on more and more samples. At the end of the training, the accuracies for both the legitimate and adversarial data sets converge to a saturated value larger than 98\%, indicating that the adversarially retrained quantum classifier is immune to the adversarial examples generated by the BIM attack. We also notice that, due to the competition between the inner maximization and outer minimization, the accuracies for the legitimate data sets for training and validation both have an oscillation at the beginning of the adversarial training process.

The above example explicitly shows that adversarial training can indeed increase the robustness of quantum classifiers against a certain type of adversarial perturbations. Yet, it is worthwhile to mention that the adversarially trained quantum classifier may only perform well on adversarial examples that are generated by the same attacking method. It does not perform as well when a different attack strategy is used by the attacker. In addition, adversarial training tends to make the quantum classifier more robust to white-box attacks than to black-box attacks due to gradient masking \cite{papernot2017practical,tramer2017space}. In fact, we expect \textit{no} universal defense strategy that is adaptive to all types of adversarial attacks,  as one approach may block one kind of attack for the quantum classifier but will inevitably leave another vulnerability open to an attacker who knows and makes use of the underlying defense mechanism.  In the field of classical adversarial learning, a novel intriguing defense mechanism that is effective against both white-box and black-box attacks has been proposed recently \cite{samangouei2018defense}. This strategy is called defense-GAN, which leverages the representative power of GAN to diminish the effect of adversarial perturbations via projecting input data onto the range of the GAN's generator before feeding it to the classifier. More recently, a quantum version of GAN (dubbed QGAN) has been theoretically proposed \cite{Lloyd2018Quantum,Demers2018Quantum} and a proof-of-principle experimental realization of QGAN has been reported with superconducting quantum circuits \cite{hu2019quantum}. Likewise, it would be interesting and important to develop a defense-QGAN strategy to enhance the robustness of quantum classifiers against adversarial perturbations. We leave this interesting topic for future study.

\section{CONCLUSION AND OUTLOOK}\label{conclusion-and-outlook}

In summary, we have systematically studied the vulnerability of quantum classifiers to adversarial examples in different scenarios. We found that, similar to classical classifiers based on deep neural networks, quantum classifiers are likewise extremely vulnerable to adversarial attacks: adding a tiny amount of carefully-crafted perturbations, which are imperceptible to human eyes or ineffective to conventional methods, into the original legitimate data (either classical or quantum mechanical) will cause the quantum classifiers to make incorrect predictions with a notably high confidence level. We introduced a generic recipe on how to generate adversarial perturbations for quantum classifiers with different attacking methods and gave three concrete examples in different adversarial settings,  including classifying real-life handwritten digit images in MNIST, simulated time-of-flight images for topological phases of matter, and quantum ground states for studying the paramagnetic/ferromagnetic quantum phase transition. In addition,
through adversarial training, we have shown that the vulnerability of quantum classifiers to
specific types of adversarial perturbations can be significantly suppressed. Our discussion is mainly focused on supervised learning based on quantum circuit classifiers, but its generalizations to the case of unsupervised learning and other types of quantum classifiers are possible and straightforward. Our results reveal a novel vulnerability aspect for quantum machine learning systems to adversarial perturbations, which would be crucial for practical applications of quantum classifiers in the realms of both artificial intelligence and machine learning phases of matter as well. 

It is worthwhile to clarify the differences between the quantum adversarial learning discussed in this paper and the quantum generative adversarial networks (QGAN) studied in previous works \cite{Lloyd2018Quantum,Demers2018Quantum,Zeng2019Learning,hu2019quantum,chakrabarti2019quantum}. A QGAN contains two major components, a generator and a discriminator, which are trained alternatively in the way of an adversarial game: at each learning round, the discriminator optimizes her strategies to identify the fake data produced by the generator, whereas the generator updates his strategies to  fool the discriminator. At the end of the training, such an adversarial procedure will end up at a Nash equilibrium point, where the generator produces data that match the statistics of the true data from the original training set and the discriminator can no longer distinguish the fake data with a probability larger than one half. The major goal of QGAN is to produce new data (either classical or quantum mechanical) that match the statistics of the training data, rather than to generate adversarial examples that are endowed with wild patterns.

This work only reveals the tip of the iceberg. Many important questions remain unexplored and deserve further investigations.  First, the existence of adversarial examples seems to be a fundamental feature of quantum machine learning applications in high-dimensional spaces \cite{liu2019vulnerability} due to the concentration of measure phenomenon \cite{Ledoux2001Concentration}.  Thus, we expect that various machine learning approaches to a variety of high-dimensional problems, such as separability-entanglement classification \cite{lu2018separability,ma2018transforming}, quantum state discrimination \cite{chefles2000quantum}, quantum Hamiltonian learning \cite{Wang2017Experimental}, and quantum state tomography \cite{Torlai2018Neural,Carrasquilla2019Reconstructing}, should also be vulnerable to adversarial attacks. Yet, in practice how to find out all possible adversarial perturbations in these scenarios and develop appropriate countermeasures feasible in experiments to strengthen the reliability  of these approaches still remain unclear. Second, in classical adversarial learning a strong ``No Free Lunch" theorem has been established recently\cite{tsipras2018robustness,fawzi2018adversarial,gilmer2018adversarial}, which shows that there exists an intrinsic tension between adversarial robustness and generalization accuracy. In the future, it would be interesting and important to prove a quantum version of such a profound theorem and study its implications in practical applications of quantum technologies. In addition, there seems to be a deep connection between the existence of adversarial perturbations in quantum deep learning and the phenomenon of orthogonality catastrophe in quantum many-body physics \cite{Anderson1967Infrared, Deng2015Exponential}, where adding a week local perturbation into a metallic or many-body localized Hamiltonian will make the ground state of the slightly-modified Hamiltonian  orthogonal to that of the original one in the thermodynamic limit. A thorough investigation of this will provide new insight into the understanding of both adversarial learning and orthogonality catastrophe. Finally, an experimental demonstration of quantum adversarial learning should be a crucial step towards practical applications of quantum technologies in artificial intelligence in the future.


\begin{acknowledgements}
We thank Nana Liu, Peter Wittek, Ignacio Cirac, Roger Colbeck, Yi Zhang, Xiaopeng Li, Christopher Monroe, Juan Carrasquilla, Peter Zoller, Rainer Blatt, John Preskill, Zico Kolter, Al{\'a}n Aspuru-Guzik, and Peter Shor for helpful discussions. S.L. would like to further thank Mucong Ding, Weikang Li, Roger Luo, and  Jin-Guo Liu for their help in developing the code for implementing the adversarial machine learning process. This work was supported by the Frontier Science Center for Quantum Information of the Ministry of Education of China, Tsinghua University Initiative Scientific Research Program, and the National key Research and Development Program of China (2016YFA0301902). D.-L. D. acknowledges in addition the support from the National Thousand-Young-Talents Program and the start-up fund from Tsinghua University (Grant No. 53330300319).
\end{acknowledgements}

\appendix
\label{appendix}

\section{Attack Algorithms}\label{app:AM}

As mentioned in the main text, the type of attacks we consider is mainly evasion attack  from the perspective of attack surface. Evasion attack is the most common type of attack in classical adversarial learning \cite{vorobeychik2018adversarial}. In this setting, the attacker attempts to deceive the classifier by adjusting malicious samples during the testing phase. This setting assumes no modification of the training data, which is in sharp contrast to poisoning attack, where the adversary tries to poison the training data by injecting carefully-crafted samples to compromise the whole learning process. Within the evasion-attack umbrella, the attacks considered in this paper can be further categorized into additive or functional, targeted or untargeted, and white-box or black-box attacks along different classification dimensions. Here, in this Appendix, we give more technique details about the attack algorithms used.

\subsection{White-box attacks}\label{app:WBA}

White-box attacks assume full information about the classifier, so the attacker can exploit the gradient of the loss function: \(\nabla_{\mathbf{x}}{L}(h(\mathbf{x}+\delta;\theta),y)\). For the convenience and conciseness of the presentation, we will use $\mathbf{x}$ ($y$) and $|\psi\rangle_{\text{in}}$ ($\mathbf{a}$) interchangeably to represent the input data (corresponding label) throughout the whole Appendix sections. Based on the information of gradients, a number of methods have been proposed in the classical adversarial learning community to generate adversarial samples. In this work, we adopt some of these methods to the quantum setting, including the FGSM, BIM, and PGD methods.  In the following, we introduce these methods one by one and provide a pseudocode for each method.

\begin{figure}
\begin{algorithm}[H]
\caption{Quantum-adopted Fast Gradient Sign Method}
\begin{algorithmic}[1]  

\Require The trained quantum classifier $h$, loss function $L$, the legitimate sample $(|\psi\rangle_{\text{in}},\mathbf{a})$.
\Require The perturbation bound $\epsilon$
\Ensure An adversarial example $\mathbf{x}^*$.
\State Input $|\psi\rangle_{\text{in}}$ into $F$ to obtain $\nabla_x L(h(|\psi\rangle;\Theta^*),\mathbf{a})$
\For {Every component $x_i$ of $|\psi\rangle_{\text{in}}$}
	\State $\delta_i=\epsilon\cdot \text{sign}(\nabla_{x_i} L(h(|\psi\rangle_{\text{in}};\Theta^*),\mathbf{a})$
	\State $x^*_i=x_i+\delta_i$
\EndFor \\
\Return $\mathbf{x}^*$ or its equivalent $|\psi\rangle^*$
\end{algorithmic}
\label{algo:QFGSM}
\end{algorithm}
\end{figure}

\textit{Quantum-adapted FGSM  method (Q-FGSM).}---The FGSM method is a simple one-step scheme for obtaining adversarial examples and has been widely used in the classical adversarial machine learning community \cite{goodfellow2014explaining,madry2017towards}. It calculates the gradient of the loss function with respect to the input of the classifier. The adversarial examples are generated using the following equation:
\begin{eqnarray}\label{eq:attack-FGSM}
\mathbf{x}^*=\mathbf{x}+\epsilon \cdot \text{sign}(\nabla_{\mathbf{x}} L(h(|\psi\rangle_{\text{in}};\Theta^*),\mathbf{a})),
\end{eqnarray}
where $L(h(|\psi\rangle_{\text{in}};\Theta^*),\mathbf{a})$ is the loss function of the trained quantum classifier, $\epsilon$ is the perturbation bound, $\nabla_{\mathbf{x}}$ denotes the gradient of the loss with respect to a legitimate sample $\mathbf{x}$ with correct label $\mathbf{a}$, and $\mathbf{x}^*$ denotes the generated adversarial example corresponding to $\mathbf{x}$. For the case of additive attacks, where we modify each component of the data vector independently, $\nabla_{\mathbf{x}}$ is computed componentwise and a normalization of the data vector will be performed if necessary. For the case of functional attacks, we use a layer of parametrized local unitaries to implement the perturbations to the input data $|\psi\rangle_{\text{in}}$. In this case, $\nabla_{\mathbf{x}}$ is implemented via the gradient of the loss with respect to the parameters defining the local unitaries. The Eq. \eqref{eq:attack-FGSM} should be understood as: 
\begin{eqnarray}
\omega^* &=& \epsilon \cdot \text{sign} (\nabla_{\omega}L(h(U(\omega)|\psi\rangle_{\text{in}};\Theta^*),\mathbf{a}),\\
|\psi\rangle_{\text{adv}} &=& U(\omega^*)|\psi\rangle_{\text{in}},
\end{eqnarray}
where $\omega$ denotes collectively all the parameters for the local unitaries. A pseudocode representation of the Q-FGSM algorithm for the case of additive attacks is shown in Algorithm \ref{algo:QFGSM}.  The pseudocode for the case of functional attacks is similar and straightforward, thus been omitted for brevity.

\textit{Quantum-adapted BIM method (Q-BIM).}---The BIM method is a straightforward extension of the basic FGSM method \cite{BIM}. It generates adversarial examples by iteratively applying the FGSM method with a small step size $\alpha$:
\begin{eqnarray}
\mathbf{x}_{k+1}^{*} = \pi_C[\mathbf{x}_{k}^{*}+\alpha \cdot \operatorname{sign}\left(\nabla_{\mathbf{x}} L\left(h(|\psi\rangle_k^*;\Theta^*), \mathbf{a}\right)\right)],
\end{eqnarray}
where $\mathbf{x}^*_k$ denotes the modified sample at step $k$ and $\pi_C$ is projection operator that normalizes the wavefunction.
A pseudocode representation of the Q-BIM algorithm for the case of additive attacks is shown in Algorithm \ref{algo:QBIM}.

\begin{figure}
\begin{algorithm}[H]
\caption{Quantum-adapted Basic Iterative Method}
\begin{algorithmic}[1]  
\Require The trained model $h$, loss function $L$, the legitimate sample $(|\psi\rangle_{\text{in}},\mathbf{a})$.
\Require The perturbation bound $\epsilon$, iteration number $T$, decay factor $\mu$, upper and lower bound $x_{\text{min}}, x_{\text{max}}$.
\Ensure An adversarial example $|\psi\rangle^*$.
\State $|\psi\rangle^*_0=|\psi\rangle_{\text{in}}$
\State $\alpha=\frac{\epsilon}{T}$
\For {$k=1,\dots,T$}
	\State Input $|\psi\rangle_{i-1}$ into $F$ to obtain $\mathbf{b}_k=\nabla_x L(h(|\psi\rangle_{k-1};\theta),\mathbf{a})$
	\For {Every component $(\mathbf{x}_k)_j$ of $|\psi\rangle^*_{k-1}$}
		\State $\delta_j=\alpha\cdot \text{sign}((\mathbf{b}_k)_j)$
		\State $(\mathbf{x}_k)_j=(\mathbf{x}_{k-1})_j+\delta_j$
	\EndFor
	\State $(\mathbf{x}_k)=\pi_C(\mathbf{x}_k)$
\EndFor \\
\Return $|\psi\rangle^*=|\psi\rangle_T$
\end{algorithmic}  
\label{algo:QBIM}
\end{algorithm}
\end{figure}

\subsection{Black-box attacks: transfer attack}\label{app:BBA}
Unlike in the white-box setting, black-box attacks assume that the adversary does not have full information about either the model or the algorithm used by the learner. In particular, the adversary does not have the information about the loss function used by the quantum classifier, thus cannot use the gradient-based attacking methods to generate adversarial examples. Yet, for simplicity we do assume that the attacker has access to a vast dataset to train a local substitute classifier that approximates the decision boundary of the target classifier. Once the substitute classifier is trained with high confidence, any white-box attack strategy can be applied on it to generate adversarial examples, which can be used to deceive the target classifier due to the transferability property of adversarial examples. In this work, we consider the transfer attack in a more exotic setting, where we use different classical classifiers as the local substitute classifier to generate adversarial examples for the quantum classifier. The two classical classifiers are based on the CNN and FNN, respectively. In Table \ref{table:MNIST-classifier}, we show the detailed structures of the CNN and FNN. To train these two classical classifiers, we use  the Adam optimizer \cite{kingma2014adam} and a batch size of $256$. The learning rate is set to be $10^{-3}$ during training. The corresponding learning process is implemented using Keras \cite{chollet2015keras}, a high-level deep learning library running on top of the TensorFlow framework \cite{abadi2016tensorflow}. After training, both the CNN and FNN classifiers achieve a remarkably high accuracy on the legitimate testing dataset ($98.9\%$ and $99.9\%$ respectively, see Table \ref{table:transferability-table} in the main text).

\begin{table}
	\caption{Model architectures for the classical neural networks. (a) The CNN architecture consists of three layers: a 2D convolution layer, an activational ReLu layer \cite{nair2010rectified}, and a fully-connected flattening layer with \(0.5\) dropout regularization. The last layer is then connected to the final softmax classifier, which outputs the probability for each possible handwritten digit. In our case, we have four categories: 1, 3, 7, 9. (b) The feedforward neural network architecture consists of fully-connected layers and dropout \cite{srivastava2014dropout} layers with a dropping rate \(0.1\), which are important for avoiding overfitting. \label{table:MNIST-classifier}}
	\begin{ruledtabular}
	\begin{tabular}{ll}
		{Classifier based on CNN} & {Classifier based on FNN}  \\
		\colrule
		Conv(64,8,8)+ReLu & FC(512)+ReLu\\
		Conv(128,4,4)+ReLu & Dropout(0.1)\\
		Conv(128,2,2)+ReLu & FC(53)+ReLu\\
		Flatten & Dropout(0.1)\\
		FC(4)+Softmax & FC(4)+Softmax\\
	\end{tabular} 
	\end{ruledtabular}
\end{table}

We use three different methods, namely the BIM, FGSM and MIM methods, to attack both the CNN and FNN classifiers in a white-box setting to obtain adversarial examples. These attacks are implemented by using of Cleverhans \cite{papernot2016technical}. For the BIM attack, the number of attack iteration is set to be ten  and the step size \(\alpha\) is set to be $0.01$. For the FGSM attack, the number of iteration is one and the step size is set to be $0.3$.  For the MIM method, the number of attack iterations is set to be ten, the step size is set to be $0.06$, and the decay factor \(\mu\) is set to be $1.0$.  A detailed description of the MIM method, together with a pseudocode, can be find in Ref.~\cite{Jiang2019Vulnerability}. The performance of both classifiers on the corresponding sets of adversarial examples is shown in Table \ref{table:transferability-table} in the main text, from which it is clear that the attack is very effective (the accuracy for both classifiers decreases to a value less than $1\%$). After the adversarial examples were generated, we test the performance of the quantum classifiers on them and find  that its accuracy decrease noticeably (see Table~\ref{table:transferability-table} in the main text).


%

\end{document}